\documentclass{emulateapj}
\usepackage{epstopdf}
\usepackage{graphicx}
\usepackage{epsfig}

\usepackage{hyperref}
\providecommand{\eprint}[1]{\href{http://arxiv.org/abs/#1}{#1}}
\providecommand{\bibinfo}[2]{\ifthenelse{\equal{#1}{isbn}}{%
\href{http://cosmologist.info/ISBN/#2}{#2}}{#2}}
\providecommand{\adsurl}[1]{\href{#1}{ADS}}

\newcommand{\avg}[1]{\ensuremath{\langle #1 \rangle}}
\newcommand{\bma}{\begin{math}}
\newcommand{\ema}{\end{math}}
\newcommand{\beq}{\begin{equation}}
\newcommand{\eeq}{\end{equation}}
\newcommand{\beqa}{\begin{eqnarray}}
\newcommand{\eeqa}{\end{eqnarray}}
\newcommand{\bc}{\begin{center}}
\newcommand{\ec}{\end{center}} 
\newcommand{\bit}{\begin{itemize}}
\newcommand{\eit}{\end{itemize}}

\begin{document}

\submitted{\today. Submitted to \apj.} 

\title{Simulations and Analytic Calculations of Bubble Growth
During Hydrogen Reionization}
\author{Oliver Zahn\altaffilmark{1}, Adam Lidz\altaffilmark{1},
Matthew McQuinn\altaffilmark{1}, Suvendra Dutta\altaffilmark{1}, Lars Hernquist\altaffilmark{1},
  Matias Zaldarriaga\altaffilmark{1,2}, Steven R. Furlanetto\altaffilmark{3}}

\email{ozahn@cfa.harvard.edu,alidz@cfa.harvard.edu}
\altaffiltext{1}{Harvard-Smithsonian Center for Astrophysics, 60 Garden Street, Cambridge, MA 02138}
\altaffiltext{2}{Jefferson Laboratory of Physics; Harvard University; Cambridge, MA 02138}
\altaffiltext{3} {Yale Center for Astronomy and Astrophysics, Yale University, \
260 Whitney Avenue, New Haven, CT 06520-8121}

\begin{abstract}
We present results from a large volume simulation of Hydrogen reionization. 
We combine 3d radiative transfer calculations and an N-body simulation, 
describing structure formation in the intergalactic medium (IGM), to
detail the growth of HII regions around high redshift galaxies.
Our N-body simulation tracks $1024^{3}$ dark matter particles, in a cubical
box of co-moving side length $L_{\rm box} = 65.6$ Mpc $h^{-1}$. This large
volume allows us to accurately characterize the size distribution of 
HII regions
throughout most of the reionization process. At the same time, our simulation 
resolves many of the small galaxies likely responsible for reionization.
It confirms a picture anticipated by analytic models: HII regions grow collectively around highly-clustered sources, and have a well-defined 
characteristic size, which evolves from a sub-Mpc scale at the
beginning of reionization to $R \gtrsim 10$ co-moving Mpc towards the
end. We show that in order to obtain this qualitative picture, source
resolution must not be sacrificed at too great a level.
We present a detailed statistical description of our results, and compare
them with a numerical hybrid scheme based on the analytic model by Furlanetto, Zaldarriaga, and Hernquist. This model associates HII regions with large-scale overdensities and is based on the excursion set formalism.  
We find that the analytic calculation reproduces the size distribution
of HII regions, the power spectrum of the ionization field, and the 21
cm power spectrum of the full radiative transfer simulation remarkably 
well.  
The ionization field from the radiative transfer simulation, however,
has more small scale structure than the analytic calculation, owing to Poisson
scatter in the simulated abundance of galaxies on small scales. We propose and 
validate a simple scheme to incorporate this scatter into our
calculations. Our results suggest that analytic calculations are sufficiently accurate
to aid in predicting and interpreting the results of future 21 cm surveys. In particular, our fast
numerical scheme is useful for forecasting constraints from future
21 cm surveys, and in constructing mock surveys to test data analysis procedures.
 \end{abstract}


\section{Introduction} \label{sec:intro}

The epoch of reionization (EOR) is a pivotal stage in the process of 
cosmological structure formation, marking the birth of the first luminous objects, a key 
landmark as the universe transforms from the relatively smooth state probed
by the cosmic microwave background (CMB), to its present day complexity.
Our current observational constraints on reionization come from 
Ly$\alpha$ forest absorption spectra towards high redshift 
quasars and constraints on the evolution of the ionizing background (e.g. \citealt{Fan:2005es}), from measurements 
of the high redshift galaxy luminosity function from narrow-band Ly$\alpha$-emission
searches \citep{Malhotra:2005qf}, and from measurements of the large
scale CMB E-mode polarization \citep{Page:2006hz, Spergel:2006hy}. 
The claimed size of HII regions surrounding individual quasars has also been used to infer limits on the neutral fraction \citep{Mesinger:2004rk,Wyithe:2004jw}. 
While valuable, each of these observational  
probes has its limitations, and some of the current constraints are relatively meager. Quasar absorption spectra are limited in part by the high Ly$\alpha$ absorption
cross section: by $z \sim 6$, even a highly ionized IGM completely absorbs quasar flux
in the Ly$\alpha$ forest.  
The constraints from
narrow-band Ly$\alpha$ searches 
are subtle to interpret (e.g. \citealt{Furlanetto:2005ir}), and
restricted
to narrow redshift windows around $z=5.7$ and $z=6.5$, where
Ly$\alpha$ falls in the observed  
optical 
band, and avoids contamination from bright sky lines (e.g. \citealt{Rhoads:2002dh}). The CMB 
polarization
measurements constrain only an integral over the ionization history,
and are potentially 
sensitive to foreground contamination \citep{Kogut:2003et,Page:2006hz}.

The study of reionization may be revolutionized by future experiments aimed at detecting 21 cm emission
from the high redshift IGM. These experiments should provide three-dimensional information
regarding the distribution of high redshift neutral hydrogen, constraining the topology of reionization, and 
its redshift evolution (e.g. \citealt{Madau:1996cs, Zaldarriaga:2003du}). Several low frequency radio
telescopes are presently ramping up to detect this signal: the Mileura Wide Field Array
(MWA) \citep{Bowman:2005cr} \footnote{http://web.haystack.mit.edu/arrays/MWA/}, the PrimeavAl Structure Telescope (PAST)
\citep{Pen:2004de}, and the Low Frequency Array (LOFAR)
\footnote{http://www.lofar.org}, while another second generation experiment, the Square Kilometer Array (SKA)\footnote{http://www.skatelescope.org/}, 
is in the planning stage. 
These measurements will be dominated by foreground contamination, but in contrast to the IGM signal, the foregrounds
are expected to be smooth in frequency, facilitating their removal \citep{Zaldarriaga:2003du}.
The 21 cm data will be supplemented
by further quasar absorption spectra (including clues from
metal absorption lines: \citealt{Oh:2002ry}, \citealt{Becker:2005vg}), high redshift gamma ray bursts
(e.g. \citealt{Barkana:2003ja,Totani:2005ng}), high redshift galaxy surveys
(e.g. \citealt{Kneib:2004dq}), and  small-scale CMB measurements \citep{Santos:2003jb,Zahn:2005fn,Mcquinn:2005ce}, providing
a wealth of observational data on the process of reionization.

Detailed theoretical modeling (see e.g.
\citealt{Gnedin:2000uj,Razoumov:2002,Ciardi:2003ia,Sokasian:2001xh, 
Sokasian:2001na,Furlanetto:2004ha,Furlanetto:2004nh,Kohler:2005gg,Iliev:2005sz,Mellema:2006pd}) is, however, 
required to constrain the topology of reionization from 
current and future observations. In particular, it is subtle to
infer quantities like the 
volume filling factor and size distribution of ionized regions, 
the correlation between ionized regions and large-scale 
over-density (i.e., does reionization proceed outside-in or
inside-out), and the nature of the  
ionizing sources, from observations. 

Unfortunately, numerical modeling of reionization is challenging, requiring
treatment of radiative transfer, preferably some 
treatment of gas dynamics, and a large dynamic range. A large dynamic range is required to
resolve the small mass galaxies which may make up the sources and sinks
of ionizing photons, while simultaneously sampling the distribution
of HII regions, which may be as large as $R \sim 20$ comoving Mpc $h^{-1}$ towards
the end of reionization \citep{Furlanetto:2005xx}.
For this reason, most calculations have been performed in
prohibitively small simulation boxes, or been entirely analytic (e.g. \citealt{Furlanetto:2004nh}, hereafter FZH04), 
although there has been very
recent progress towards large volume reionization simulations
\citep{Kohler:2005gg,Iliev:2005sz}. 
Indeed a skeptic might posit that, given the difficulty of simulating
reionization, we will observe the 21 cm signal before we can predict it.

In this paper, we push forward by running a large volume radiative transfer simulation.
Our work represents progress
on several fronts. First, we simulate reionization in a larger volume than most 
previous works (although see \citealt{Kohler:2005gg,Iliev:2005sz}), while
maintaining high mass resolution. This allows us to reliably calculate the size distribution
of HII regions as well as power spectra of ionization and 21 cm fields, impossible with
previous small volume simulations.
Second, we compare our results with analytic calculations based on FZH04. These models are now widely 
used, and while elegant and inspired by
previous small volume reionization simulations 
\citep{Sokasian:2003au,Sokasian:2003gf}, they remain 
untested. Our comparison also gauges the level of theoretical control in our
modeling of reionization -- i.e., how robust are our conclusions to
the details of our modeling? One  
convincing way to dissuade the above-mentioned skeptic is to demonstrate
that we can understand the gross features of our radiative transfer simulations
\emph{analytically}.    
Additionally, if analytic models are sufficiently accurate then they are
useful tools to forecast constraints  
from future experiments, and to construct mock surveys, providing
important tests of data analysis procedures.  This is important given our ignorance
of the nature of the ionizing sources: we would like to cover a large parameter
space in the source properties, prohibitive with time-consuming radiative
transfer simulations. Furthermore, future surveys will span volumes of several
cubic Giga-parsecs, a challenging task for detailed simulations.

We emphasize that our present work is only a first step towards more realistic simulations of Hydrogen reionization.
As we describe subsequently, our radiative transfer simulations miss potentially important aspects of the physics of reionization. Specifically, we include only a crude prescription for the sources of ionizing photons,
our coarse resolution underestimates the importance of recombinations -- especially if 
mini-halos are present
during reionization \citep{Haiman:2000pd,Barkana:2002sp,Shapiro:2003gx} -- and
misses small galaxies that may contribute ionizing photons, and 
we ignore feedback effects entirely. We intend to model some of these
effects in the near future \citep{Mcquinn:2006rt}).

Our work has overlap with the recent simulation and analysis of \citet{Iliev:2005sz}.
In comparison to these authors, reionization finishes significantly later in our simulation, 
near $z \sim 6.5$, as compared to $z \sim 12$, a consequence of our more conservative
prescription for the ionizing sources.  Moreover, our main present emphasis is in comparing
our radiative transfer simulation results with `hybrid simulations' based on analytic models.

The layout of this paper is as follows. In \S \ref{sec:sims} we
describe our N-body simulation, source 
prescription and radiative transfer calculation. In \S
\ref{sec:analytic} we describe our `analytic model simulation', which is more precisely an implementation of a model based on FZH04 into the cosmological realization used for the radiative transfer simulation. We will sometimes refer to this scheme losely as an `analytic calculation' although the implementation of the model is entirely numerical.
In \S
\ref{sec:stats} we present a detailed  
statistical description of our radiative transfer and analytic
results. We describe a numerical scheme that incorporates the
stochasticity of the source distribution into our analytic
calculations in \S \ref{sec:improved_mc}. We also show that if
extremely bright and rare sources reionize the IGM, bubble growth is
less collective than in our fiducial model. 

In \S \ref{sec:21cm} we compare radiative transfer and analytic model predictions for the 21 cm signal.
We conclude in \S \ref{sec:conclusion}, mentioning future research directions and emphasizing possible
improvements to our simulations.

Throughout we assume a flat, $\Lambda$CDM cosmology parameterized by: $\Omega_m=0.3$,
$\Omega_\Lambda=0.7$, $\Omega_b=0.04$, $H_0 = 100 h$ km/s/Mpc with $h=0.7$, and
a scale-invariant primordial power spectrum with $n=1$, normalized to $\sigma_8(z=0)=0.9$ \footnote{This value for $\sigma_8$ is slightly different than the value preferred by the WMAP satellite alone of $0.76\pm 0.05$ \citep{Spergel:2006hy}. When combined with other datasets, such as the Lyman-$\alpha$ forest \citep{Lewis:2006ma,Seljak:2006bg}, and weak lensing (see e.g. section 4.1.7 and Table 6 of \citealt{Spergel:2006hy}), a higher value of $\sigma_8$ can be found. Furthermore, changes in the fluctuation amplitude within the present experimental boundaries can be incorporated into our analysis by adjusting slightly the ionization efficiency parameter. This does not qualitatively affect our results, as we confirmed within the analytic scenario.}

\section{Simulations} \label{sec:sims}

We begin by running a large N-body simulation to locate dark matter halos, and
produce a cosmological density field. Next, we populate the dark matter
halos with ionizing sources, using a simple prescription
to connect mass and light (\S \ref{sec:sources}). In a subsequent 
post-processing step, we
perform a radiative transfer calculation, casting rays of ionizing
photons from our sources through the cosmological density field (\S
\ref{sec:rtsim}). We make two  
approximations with this approach. First, we assume that the gas distribution
perfectly traces the dark matter distribution, as characterized by our N-body
simulation. Second, we neglect the interplay between gas dynamics and 
radiation transport -- i.e, in reality, structure formation responds to the passage
of ionization fronts, and gas motions in turn influence the propagation
of the fronts. These effects are essential in calculating the detailed
small-scale behavior of  
ionization fronts, as fronts slow down upon impacting dense clumps
\citep{Shapiro:2003gx}, but are less 
important for our goal of capturing the large-scale size distribution of HII regions.

\subsection{N-body simulations}
\label{sec:nbody}

As noted in the introduction, we require a cosmological simulation
with a large dynamic range, in order 
to adequately sample the distribution of HII regions, while
simultaneously resolving small galaxies. 
Ideally, we
would resolve halos with virial
temperatures of $T_{\rm vir} \gtrsim 10^{4}$ K -- corresponding to a dark
matter halo mass of $M_{\rm dm} \sim 10^{8} M_\odot$ at $z \sim 6$ -- above 
which atomic line cooling is efficient.
In halos more massive than this, gas can cool, condense to form stars, and produce 
ionizing photons. This `cooling mass' therefore represents a plausible
guess as to the minimum 
host halo mass for ionizing sources.
If molecular Hydrogen cooling is efficient despite radiative feedback, however, even smaller
mass halos should host sources \citep{Haiman:1996rc}. Presently, we ignore this possibility. 
Additionally, high resolution is required to capture the clumpiness of the IGM, and properly
account for recombinations during reionization.
On the other
hand, HII regions may be larger 
than $R \gtrsim 20$ Mpc $h^{-1}$ at the end of reionization
\citep{Furlanetto:2005xx}, necessitating a large 
volume simulation. Unfortunately, to
resolve a $10^{8} M_{\rm \odot}$ halo with $32$ particles, in a simulation box of side-length $L=100$ 
Mpc $h^{-1}$, for example, requires a prohibitively large number of particles, $N_p \sim 3360^3$!

Our present N-body simulation is meant to represent a compromise between
these competing requirements of large volume, and high mass 
resolution. Specifically, our N-body simulation follows $1024^{3}$ dark matter
particles in a box of side-length, $L=65.6$ Mpc $h^{-1}$, using an enhanced
version of the TreePM code, Gadget-2 \citep{Springel:2005mi}. We run the simulation
assuming the flat LCDM cosmology specified in the introduction, with
initial conditions generated using the \citet{Eisenstein:1997jh} transfer 
function.   

\begin{figure}
\bc
\includegraphics[width=9.2cm]{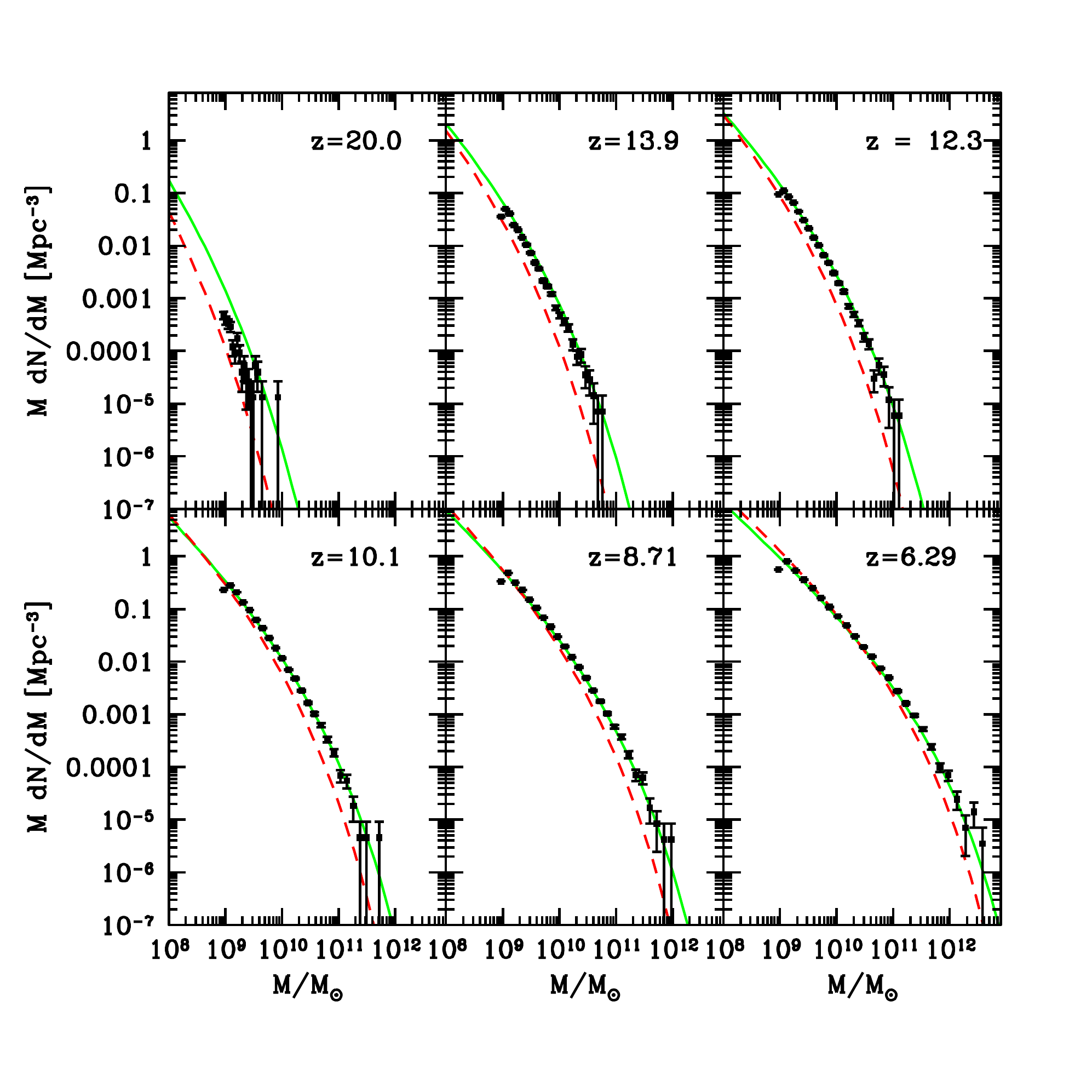}
\caption{Halo mass function from our N-body simulation. The black
points with (Poisson) error bars indicate the halo mass function from
our simulation as a function of redshift. The green curve is the Sheth-Tormen
fitting function for the halo mass function, while the red dashed line shows
the Press-Schechter fitting function.}
\label{fig:mass_func}
\ec
\end{figure}

Dark matter halos are identified from simulation snapshots, using a friends-of-friends 
algorithm (e.g., \citealt{Davis:1985rj}). Specifically, particles are
grouped into halos using a linking 
length of $b = 0.2$ times the mean interparticle separation. Linked groups
of greater than $32$ particles are considered resolved, and to constitute dark matter halos. 
This corresponds to a minimum halo mass of $10^9 M_\odot$, just an order
of magnitude above the cooling mass. 

The resulting mass function is shown in Figure \ref{fig:mass_func}, spanning
a broad redshift range between $z \sim 6 - 20$.
\citet{Barkana:2003qk} showed that if a simulation is normalized
   to the cosmic mean density, the halo mass functions will be biased. According to Figure 3 of \citet{Barkana:2003qk}, the bias introduced in our calculations should only be of order 0.1 \%.
The halo mass function is 
sampled with large dynamic range, roughly
three orders of magnitude near $z \sim 6$. The simulated mass function is 
always larger than predicted by
the Press-Schechter formalism \citep{Press:1973iz}, but generally in good agreement
with the Sheth-Tormen \citep{Sheth:1999mn} fitting formula. At the highest redshifts 
sampled, 
however, our results fall in between the two fitting formula. This is
in qualitative agreement with recent measurements from \citet{Reed:2003hp},
and \citet{Heitmann:2006eu}, although our mass function appears systematically
higher than that of \citet{Iliev:2005sz}. The figure shows that the abundance
of our lowest mass halos is systematically below theoretical expectations,
likely a consequence of our limited mass resolution. 
As a conservative measure, we therefore place ionizing sources only
in halos of mass larger than $M_{\rm min} = 2 \times 10^9 M_\odot$, corresponding to 
a $64-$particle halo. 

\subsection{Ionizing Sources}
\label{sec:sources}

Our next step is to connect mass with light -- that is, we wish to populate
the dark matter halos from our N-body simulation with ionizing sources. In 
this paper, we will
adopt a very crude prescription for our ionizing sources, leaving
a more sophisticated prescription to future work. This will
facilitate comparison with the analytic models (see \S
\ref{sec:analytic}). Specifically, we populate 
each dark matter halo with a single source whose luminosity
in Hydrogen ionizing photons is directly proportional to the host halo mass,
$\dot{N} = c M_{\rm halo}$. Clearly the parameter $c$ encodes a good deal
of complicated physics, involving the efficiency of star formation,
the efficiency of producing ionizing photons, the fraction of ionizing
photons that escape from the host halo, etc.
With this single simplifying assumption, the cumulative number of ionizing
photons released by the sources, per hydrogen atom in the IGM, at time $t$ is 
$N_{\rm ph}/N_H \propto \int_0^t dt^\prime f_{\rm coll}(t^\prime)$. Here
$f_{\rm coll}(t^\prime)$ is the fraction of mass in halos with mass
$M \geq M_{\rm min} = 2 \times 10^9 M_\odot$. Using the \citet{Sheth:1999mn}
mass function, which closely matches our simulation results
(Figure \ref{fig:mass_func}), we find that
$c = 3.1 \times 10^{41}$ photons/sec/$M_\odot$ yields one photon per hydrogen atom
at $z=6.5$. (See Figure \ref{fig:xi} and associated text for a discussion).
This choice of $c$ corresponds roughly, for example, to Pop II stars, forming with an efficiency
of $f_\star = 0.1$ from a Salpeter IMF, with a stellar lifetime of $\Delta t \sim 5 \times 10^7$ yrs,
and a modest escape fraction of $f_{\rm esc} \sim 0.01$ \citep{Loeb:2004zs}. We adopt this conversion in all
subsequent calculations.

\subsection{Radiative Transfer}
\label{sec:rtsim}

We next form a coarse density field for many snapshots, spaced in 
equal time intervals of $\Delta t = 5 \times 10^7$ years and spanning a
broad redshift range from $z \sim 6 - 16$, by gridding our dark 
matter particles onto a uniform, Cartesian grid with $256^3$ mesh points. 
Our sources (\S \ref{sec:nbody} and \S \ref{sec:sources}) are tabulated at the same
time-sampling, and moved close to
the center of their corresponding cell. Occasionally, several sources land in a single
cell and our considered to be a sole, more luminous source. At $z \sim 6.5$,
near our assumed completion of reionization, there are $\simeq
330,000$ ionizing sources in our  
simulation.

With the ionizing sources and cosmological density field in hand, we
trace rays of ionizing photons through the simulation box using the adaptive ray-tracing
scheme of \citet{Abel:2001qs}, and the
code of Sokasian et al. \citep{Sokasian:2001na, Sokasian:2003au}.  Further improvements to the code were made \citep{Mcquinn:2006rt}. We refer the reader to these papers for the details of the code used here, but give a brief summary.
In short, the code assumes a sharp ionization front, and tracks the position of the front
by casting rays and integrating over the ionization front jump
condition \citep{Abel:1998qq}.  
The jump condition amounts to tabulating the number of
photoionizations and recombinations along a ray, 
halting the ray when its photon supply is exhausted.
Each source is considered separately, although the order in which
sources are processed is randomized 
at each timestep to avoid artifacts \citep{Sokasian:2003au}. 

Behind the ionization front, each source hitting a given cell
contributes a photoionization rate of  
$\Gamma_{\rm HI, s} = \bar{\sigma} \dot{N}/(4 \pi r^2_s)$,
i.e. assuming optically thin conditions  
within the front. Here $r_s$ is the distance from the
cell in question to a source, $\dot{N}$ is the number of Hydrogen
photons per second from a source, 
and $\bar{\sigma}$ is a frequency-averaged cross section, computed here assuming each source
has a spectrum $\propto \nu^{-4}$ \citep{Sokasian:2001na}. Within the
front, ionization fractions 
are computed assuming ionization equilibrium and a uniform temperature of $T=10^4 K$, and neglecting sub-grid clumping. 
We follow the approach of \citep{Sokasian:2001na} in using case B recombination rates when casting rays through the grid, while using case A coefficients to compute the ionization fraction in previously ionized cells. 
Helium is assumed to be at most singly-ionized by our soft sources, and we assume that the HeII front precisely tracks the HII front. Similarly, inside the front we assume that the 
ionized Helium (HeII) fraction 
traces the HII fraction \citep{Sokasian:2003au}. 
Note that all of these assumptions impact mainly the detailed ionization fractions within the front,
and are less important for tracking the overall size distribution of HII regions.
In contrast to
\citet{Sokasian:2001na, Sokasian:2003au},
we do not include a diffuse background
radiation field, simply allowing rays  
to wrap around the periodic box. 

Our assumption of a sharp ionizing front is justified
given the short mean free path of Hydrogen ionizing photons in the pre-reionization IGM. 
\citet{Mellema:2005ht}
present explicit comparisons between `ionization front tracking' and
more detailed calculations that 
self consistently solve for the optical depth, ionization fraction,
and temperature. At least in the 
case of a single source (their Figure 16), ionization front tracking
reproduces very closely the 
results of more detailed calculations, further justifying our approach.

\begin{figure}
\bc
\includegraphics[width=9.7cm]{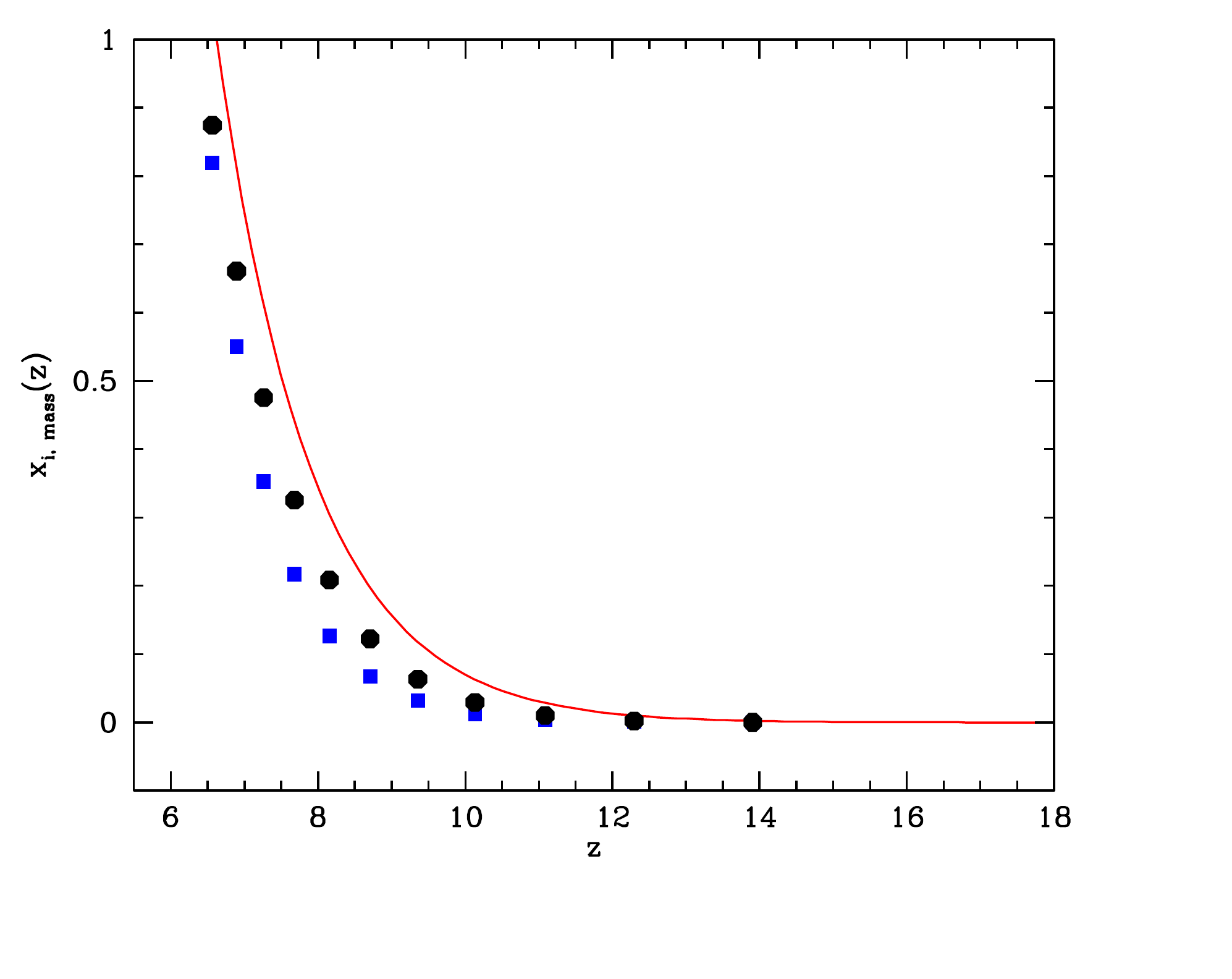}
\caption{Ionization fraction as a function of
  redshift. The black circles show the mass-weighted ionization
  fraction from the simulation, while 
the blue squares show the volume-weighted ionization fraction. The red
line is the cumulative number of ionizing photons per hydrogen atom expected 
for our ionizing sources. The close resemblance between the number of photons
per atom and the measured ionization fractions owes to the poor resolution
of our radiative transfer calculation, which underestimates the importance
of recombinations.}
\label{fig:xi}
\ec
\end{figure}

In Figure \ref{fig:xi} we plot the redshift evolution of the ionization
fraction in our simulation. The black circles show the mass-weighted
ionization fraction, while the blue squares show the volume-weighted
ionization fraction. The mass-weighted ionization fraction is somewhat
larger than the volume-weighted 
ionization fraction. This is because the ionizing sources in our
simulation are highly biased, 
and ionize their overdense environs before breaking-free to ionize neighboring 
voids (e.g. \citealt{Sokasian:2003au,Iliev:2005sz}).
The reionization process takes a fairly significant stretch of cosmic
time, with the mass-weighted ionization fraction at the level of
$x_{\rm i, m} \sim 0.1$ at $z \sim 9$, and attaining $x_{\rm i, m}  \sim 1$
only by $z \sim 6.5$.  

The evolution of the neutral fraction in our model is consistent
with, although not required 
by the measurements of e.g. \citet{Fan:2005es}, which demand only that the IGM
reionize sometime before $z \gtrsim 6$. 
Our model produces an electron scattering optical
depth of $\tau_e = 0.06$, on the low side of 
CMB constraints \citep{Page:2006hz}, which
suggest $\tau_e = 0.09 \pm 0.03$. We emphasize that our choice of $c$ (\S \ref{sec:sources}) was
calibrated so that reionization ends slightly above $z \gtrsim 6$, so this should be viewed as a consequence of our assumptions, rather than a theoretical
prediction. Although our model is tuned to give late reionization, analytic models
find that the size distribution of HII regions depends primarily on the bias of the
ionizing sources, with only an implicit dependence on redshift \citep{Furlanetto:2005ax}. The size distribution
of HII regions at a given ionization fraction is therefore expected to be a robust
result, independent of our detailed assumptions about the efficiency of the ionizing sources, (although see \citealt{Furlanetto:2005ax}, \S \ref{sec:improved_mc} for caveats).

Note that our simulation terminates slightly before reionization
completes ($x_i(z) \sim 1$). 
We stop our calculation early because we do not include a `diffuse
background' in our simulation \citep{Sokasian:2001xh}, and so our
calculation becomes very expensive at 
the end of reionization when rays wrap around the simulation box
several times. In any event, the 
tail end of reionization is likely poorly modeled in our simulation, since this stage may be 
regulated primarily by Lyman limit systems \citep{Miralda-Escude:2000,Furlanetto:2005xx}, which are missing in our analysis.

In our simulation, the mass-weighted ionization fraction closely
tracks the cumulative number of ionizing 
photons per Hydrogen atom emitted by our ionizing sources (see the solid red line in
Figure \ref{fig:xi}), but this is
partly an artifact of the  
poor resolution of our radiative
transfer calculation. Our low grid resolution
underestimates the amount of small scale structure in the density
field, and hence the importance 
of recombinations. In the future, we intend to model recombinations as
`subgrid physics', accounting for  
enhancements in the recombination rate owing
to unresolved small scale structure (see
e.g. \citet{Kohler:2005gg}). Presently, we caution that 
we are under-estimating the number of ionizing photons per Hydrogen
atom required to complete reionization. 
Furthermore, we expect reionization to be even more extended than in
our calculation, since recombinations 
should slow the growth of HII regions.

\section{Numerical scheme based on analytic considerations}
\label{sec:analytic}

As motivated in the introduction, we compare our results with a hybrid scheme inspired by the analytic model of FZH04. In this section, we describe this hybrid model, but also refer the reader to FZH04, \citet{Zahn:2005fn} for more information. The major advantage of our
implementation over a purely analytic calculation is that the hybrid scheme, which amounts to a Monte-Carlo realization of the analytic model, can 
capture the asphericity of HII regions during reionization. 

The hybrid scheme starts by considering spheres of varying radius surrounding 
every point in the IGM. Within
each such sphere, we calculate the total ionizing photon yield of the sources, and the total enclosed mass in neutral Hydrogen. 
In the event that the photon yield in a sphere around
a given point exceeds the number of interior Hydrogen atoms, the point is considered ionized. 
FZH04 show that this amounts to a barrier-crossing problem, solvable with tools from 
the excursion set formalism \citep{Bond:1990iw}.
Specifically, we assume that the mass contained in halos in a region of total mass m, and 
over-density $\delta_m$,
follows the extended Press-Schechter formula for the collapse fraction (e.g. \citealt{Lacey:1993}):
\beq
f_{\rm coll.}(m \geq m_{\rm min}|\delta_m,z) ={\rm erfc} \left[ \frac{\delta_c(z)-\delta_m}{\sqrt 2
    [\sigma^2_{\rm min}-\sigma^2(m)]}\right] .
\label{eq:fcoll}
\eeq
Here $\sigma^2(m)$ is the (present day) linear variance of density
fluctuations on the scale $m$, $\delta_c(z)=1.686/D(z)$ is the
critical density for collapse scaled to today, and $D(z)$ is the linear growth factor. The 
quantity
$\sigma^2_{\rm min}$ is the linear variance smoothed on a mass scale corresponding to that of the minimum mass 
halo that can host ionizing sources, presently $m_{\rm min}=2 \times 10^9 M_\odot$ (\S \ref{sec:nbody}).

For constant mass to light sources, the criterion for a region to self-ionize is then:
\beq
\alpha \int_0^t dt^\prime f_{\rm coll.} (m \geq m_{\rm min}|\delta_m,t^\prime) \geq 1 \, ,
\label{eq:barrier}
\eeq
where $\alpha$ is an efficiency factor linking halo mass and ionizing photon yield. 
A region can self-ionize if it is sufficiently overdense to satisfy
the inequality in Equation (\ref{eq:barrier}).
Note that this is
a slight modification from FZH04 to the case of ionizing sources with a constant mass to light ratio,
as assumed in our simulation. In practice, however, we find that the threshold criterion of 
Equation (\ref{eq:barrier}) gives quantitatively similar results to that of FZH04, although it produces
slightly larger HII regions.

Our hybrid scheme then amounts to smoothing the linear density field generated from the initial
conditions of our N-body simulation, and checking whether cells satisfy the condition of Equation (\ref{eq:barrier}).
Algorithmically, we start by considering large spheres (comparable to the size of our simulation box),
and gradually stepping down in radius, eventually reaching smoothing scales comparable to that
of our simulation pixels. At each radius we keep track of which cells satisfy the 
condition of Equation (\ref{eq:barrier}). In the event that a cell does not cross the barrier, i.e. satisfy
the condition of Equation (\ref{eq:barrier}), \emph{at any smoothing scale}, the cell is considered neutral. 
Proceeding from large smoothing scales and progressing downward to smaller smoothing scales 
ensures that we account correctly for 
cells that are ionized by neighboring sources (FZH04). 
At this stage, we have a map of the ionization field consisting of `1's (completely ionized pixels), and
`0's (completely neutral pixels). This is a good approximation to the true equilibrium ionization fractions,
which are expected to be very close to unity. 

This scheme is quite fast: for our present
$256^3$ grid calculation, with $50$ logarithmic smoothing steps, the computation (at a given redshift) takes only $\sim 12$ minutes
on a desktop computer with a 3 GHz processor. This is vastly 
more efficient than our full radiative
transfer calculation: our N-body simulation
takes 38 hours to run down to $z \sim 6$ using $134$ 2 GHz processors, and our post-processing calculation requires a few additional days of running time on a large memory computer.
With our rapid numerical scheme, we can produce an ionization map based on the analytic model and compare with our radiative transfer simulations. Using precisely the initial conditions from our
N-body simulation in our hybrid calculation allows us to compare radiative transfer and analytic ionization fields on a cell-by-cell basis. 

Before presenting this comparison, there are a few more pertinent technical details. 
Ideally, we would compare the analytic and radiative transfer calculations
with identical assumptions regarding the ionizing efficiency of our sources, i.e. we should
calibrate $\alpha$ in Equation (\ref{eq:barrier}) based on the source prescription of \S \ref{sec:sources}.
In practice there are several difficulties with matching precisely the simulated source prescription. 
Most important, Equation (\ref{eq:fcoll}) is derived assuming sharp $k$-space filtering, while  
our smoothing procedure adopts a spherical top-hat in \emph{real space}. This slight inconsistency in our 
modeling means that our model does not conserve photons precisely, affecting the ionization fraction
for a given source efficiency, $\alpha$ (see the Appendix). Further, our simulated
mass function is closer to the Sheth-Tormen fitting formula \citep{Sheth:1999mn} than the Press-Schechter
\citep{Press:1973iz} mass function, and we require an analogue of Equation 
(\ref{eq:fcoll}) for the Sheth-Tormen
mass function \citep{Barkana:2003qk,Furlanetto:2005ax}. We improve on some of these shortcomings in \S \ref{sec:improved_mc}.
The upshot of this is that, in order to compare with our radiative transfer simulations,
we adjust $\alpha$ in Equation (\ref{eq:barrier}) at each redshift to match the (volume-weighted)
ionization fraction. This readjustment is usually of order $20\%$. 

\begin{figure}
\bc
\includegraphics[width=8.5cm]{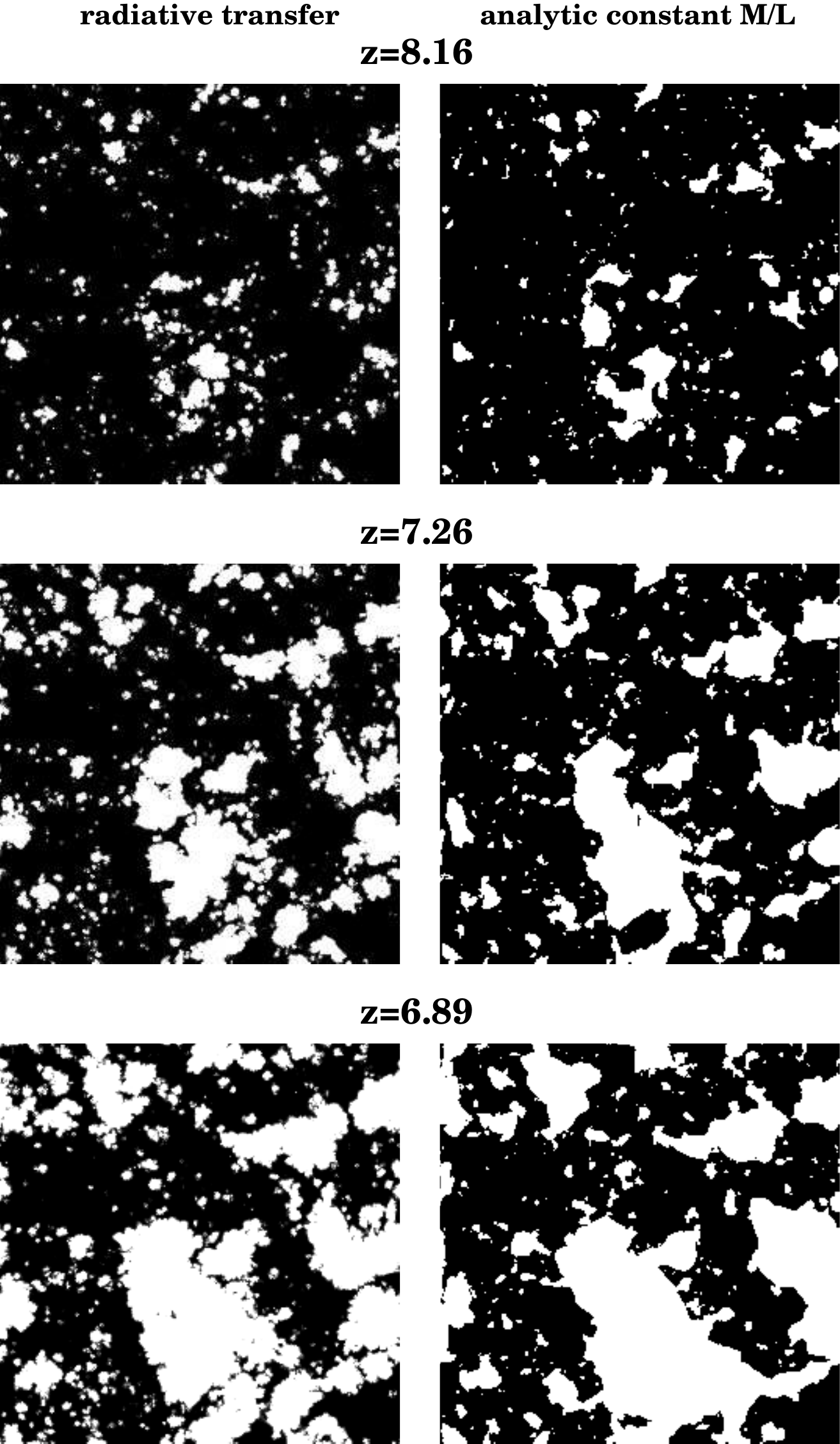}
\caption{Maps of the ionization field. The left column shows HII regions for a thin slice
  through our radiative transfer simulations at redshifts
  z=8.16, z=7.26, and z=6.89 (top to bottom). The volume-weighted ionization fraction
  at these redshifts is  $x_{\rm i, v} = 0.11, 0.33$ and $0.52$, respectively.
  The slices are $0.25$ Mpc/$h$ deep, and 
  $65.6$ Mpc/$h$ on a side. The right panel shows the same using our hybrid simulation scheme, as applied to the initial conditions used in our radiative transfer simulation. The
  analytic modeling agrees well with the more detailed simulation, although there is more small
  scale structure in the map from the radiative transfer simulation (see text).}
\label{fig:maps.xx}
\ec
\end{figure}

We show examples of the resulting ionization maps in
Figure \ref{fig:maps.xx}. The \emph{left column} shows 
thin slices through the radiative transfer simulation at three
different stages in the reionization 
process: $z = 8.16$, $7.26$ and $6.89$ when the (volume-weighted)
ionization fraction is $x_{\rm i, v} = 0.11, 
0.33$, and $0.52$ respectively. The \emph{right column} shows
corresponding slices from the hybrid simulation scheme. Several conclusions are immediately apparent.

First, the
ionized regions are quite 
large at the intermediate and late stages of reionization. The
ionizing sources are highly clustered, 
and HII regions quickly start growing collectively around the sources,
rapidly reaching much larger 
sizes than can be achieved by individual sources (FZH04) \footnote{The size of an HII region belonging to an individual average source would be too small to display in this Figure, e.g. for a $10^{10}$ $M_\odot$ source shining for one simulation time step of $5 \cdot 10^7$ years this size would be only 0.3 Mpc/h.}. Second, the
hybrid simulation is in 
good general agreement with the radiative transfer simulation. The hybrid scheme seems to `locate'
the HII regions found in the radiative transfer calculation, and
additionally reproduces their 
general morphology. Third, the HII regions in the analytic calculation are a bit more `connected' 
than those in the radiative transfer simulation. Equivalently, the ionization field in
the radiative transfer simulation appears to have more
small scale structure than the ionization field from the hybrid scheme. In the following 
sections, we will quantify the visual comparison of Figure \ref{fig:maps.xx},
diagnose differences found, and refine our numerical scheme.
We contrast the morphology seen here with that from \citet{Iliev:2005sz}
in \S \ref{sec:improved_mc}.

\section{Statistical Description}
\label{sec:stats}

In this section we present a detailed statistical description of our
results. Throughout we will compare with our hybrid scheme rather than the purely analytic calculations for two reasons. First, there are technical difficulties in the analytic calculations  at intermediate ionization fractions \citep{Mcquinn:2005ce}, and second, we would like to be able to model non-spherical bubble shapes.

\subsection{The Bubble PDF}
\label{sec:bubbles}

The first statistic we consider is the probability distribution of bubble
sizes. That is, we calculate how large the HII regions are at
different stages of reionization.  
This depends somewhat on how one chooses to define contiguous
ionized volumes -- Figure \ref{fig:maps.xx} clearly illustrates that the ionized 
regions are not spherical, particularly at the end of
reionization. The ionized regions do, however,  
obtain a reasonably well-defined characteristic
size at each redshift. In order to quantify this, we require a convenient and well-motivated 
definition of `bubble' 
that we can apply consistently to the radiative transfer simulation and the hybrid scheme.

Here we adopt a definition of bubble size inspired by the excursion set formalism,
upon which our analytic calculation is based (see \citealt{Iliev:2005sz} for an 
alternate approach). Specifically, we `draw' spheres
around each point in our simulation box of varying radius, $R$, and average (smooth) the 
ionization field within each such sphere. We start by 
considering large spheres,
of volume comparable to that of our simulation box, and step downward in
size until we eventually get to the size of our simulation pixels. At each
smoothing radius, $R$, we compare the average ionization in each sphere to
a threshold ionization, $x_{\rm th}$.
A pixel is marked as `ionized'
and belonging to a bubble of radius $R$, when $R$ is 
the \emph{largest smoothing radius}
at which the pixel's smoothed ionization exceeds the 
threshold ionization, $x_{\rm th}$. 
If a given pixel fails to exceed the threshold ionization at all
smoothing scales, it is considered neutral (not ionized). 

\begin{figure}
\bc
\includegraphics[width=9.2cm]{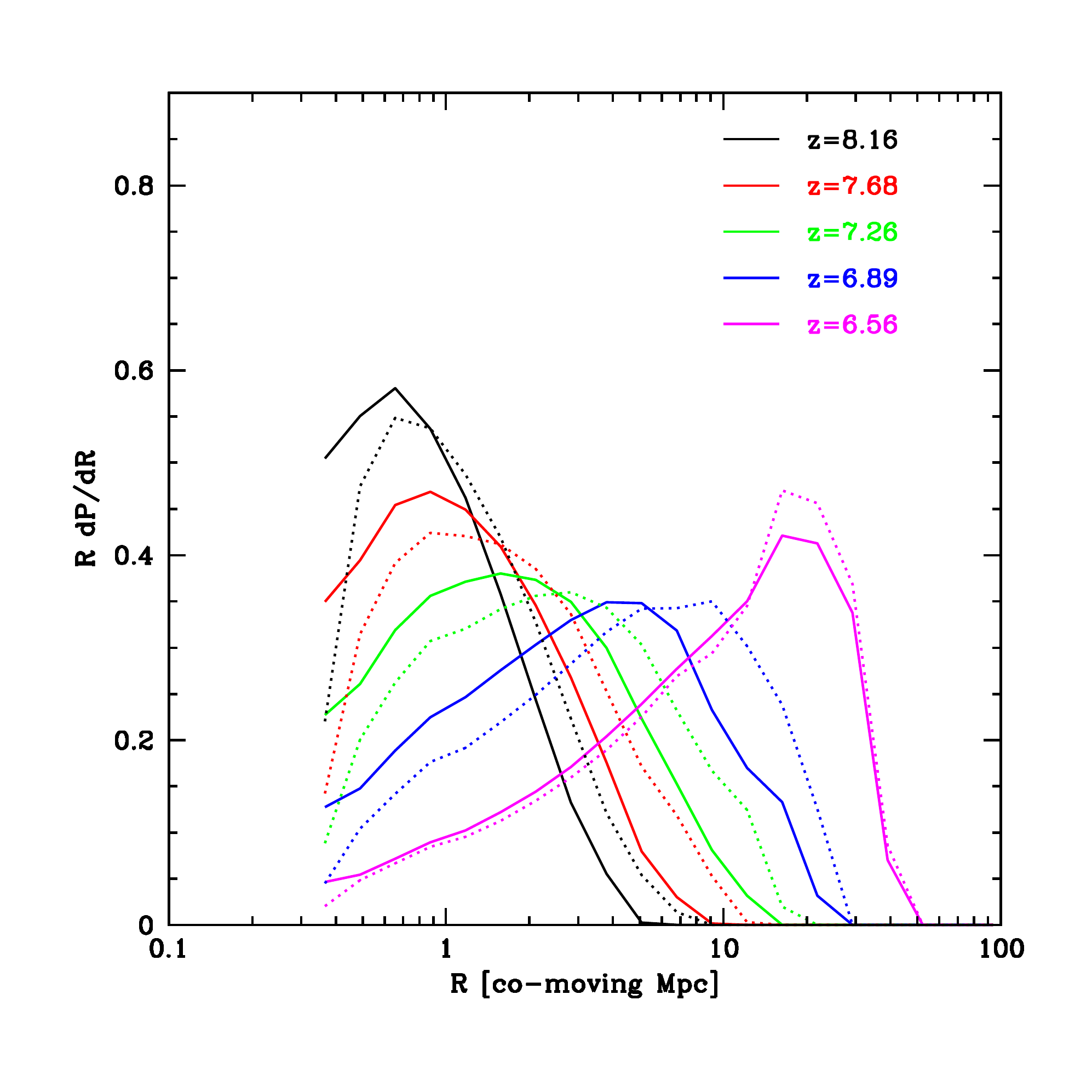}
\caption{Size distribution of HII regions as a function of
redshift. The solid curves show results from the radiative transfer simulation,
while the dotted curves are from the analytic calculation. We adopt a 
threshold ionization of $x_{\rm th} = 0.9$ (see text). The
volume-weighted ionization 
fractions at the redshifts shown are $x_{\rm i, v} = 0.11, 0.20, 0.33, 0.52, 0.77$ at
$z = 8.16, 7.68, 7.26, 6.89$ and $z = 6.56$ respectively.}
\label{fig:bubble_pdf}
\ec
\end{figure}

The bubble pdf is then derived by tabulating the fraction of ionized
pixels that lie within bubbles with radius between $R$ and $R + dR$.
With this convention, the bubble pdf is normalized to unity rather
than to the mean ionization fraction. The results of this calculation are
shown in Figure \ref{fig:bubble_pdf}, for an ionization threshold of
$x_{\rm th}=0.9$. The figure illustrates quantitatively the visual impression of
Figure \ref{fig:maps.xx} : the HII regions have a
well-defined characteristic  
size at each stage of reionization, and this characteristic scale evolves as bubbles 
around neighboring sources overlap and grow
collectively (FZH04, \citealt{Furlanetto:2005ax}). The characteristic
scale evolves from sub-Mpc scales at 
$z=8.16$, when the volume-weighted ionization fraction is 
$x_{\rm i,v}=0.11$ to $R \gtrsim 10$ Mpc co-moving at $z=6.56$ when
the volume-weighted ionization fraction is $x_{\rm i, v}=0.77$. 
The large size of HII regions at high ionization fraction implies that large
volume simulations are required to adequately sample this stage of
reionization (\citealt{Barkana:2003qk}, FZH04, \citealt{Iliev:2005sz}).
The precise value of the characteristic bubble size depends somewhat on the number
we adopt for the threshold ionization. For instance, if we instead adopt
the less stringent threshold of $x_{\rm th}=0.7$, the characteristic size increases
by a factor of $\sim 2$ near $z = 8.16$. Again, while our definition
of bubble-size is somewhat arbitrary, the bubbles nevertheless have a well-defined characteristic scale \citep{Furlanetto:2005ax}, and 
our algorithm can be applied consistently to each of the analytic model and radiative transfer ionization maps.

The dotted lines indicate that our hybrid scheme reproduces the
 bubble pdf simulated through radiative transfer quite accurately, roughly matching the characteristic bubble size and its trend with redshift.  
The hybrid scheme however leads to slightly larger HII regions at
all but the final redshift. We will discuss this difference in future sections. At the final redshift, the agreement is almost exact, however here our simulated volume is too small to provide a representative sample.

\subsection{Power Spectra of the ionized fraction}
\label{sec:pspec}
\begin{figure}
\bc
\includegraphics[width=9cm]{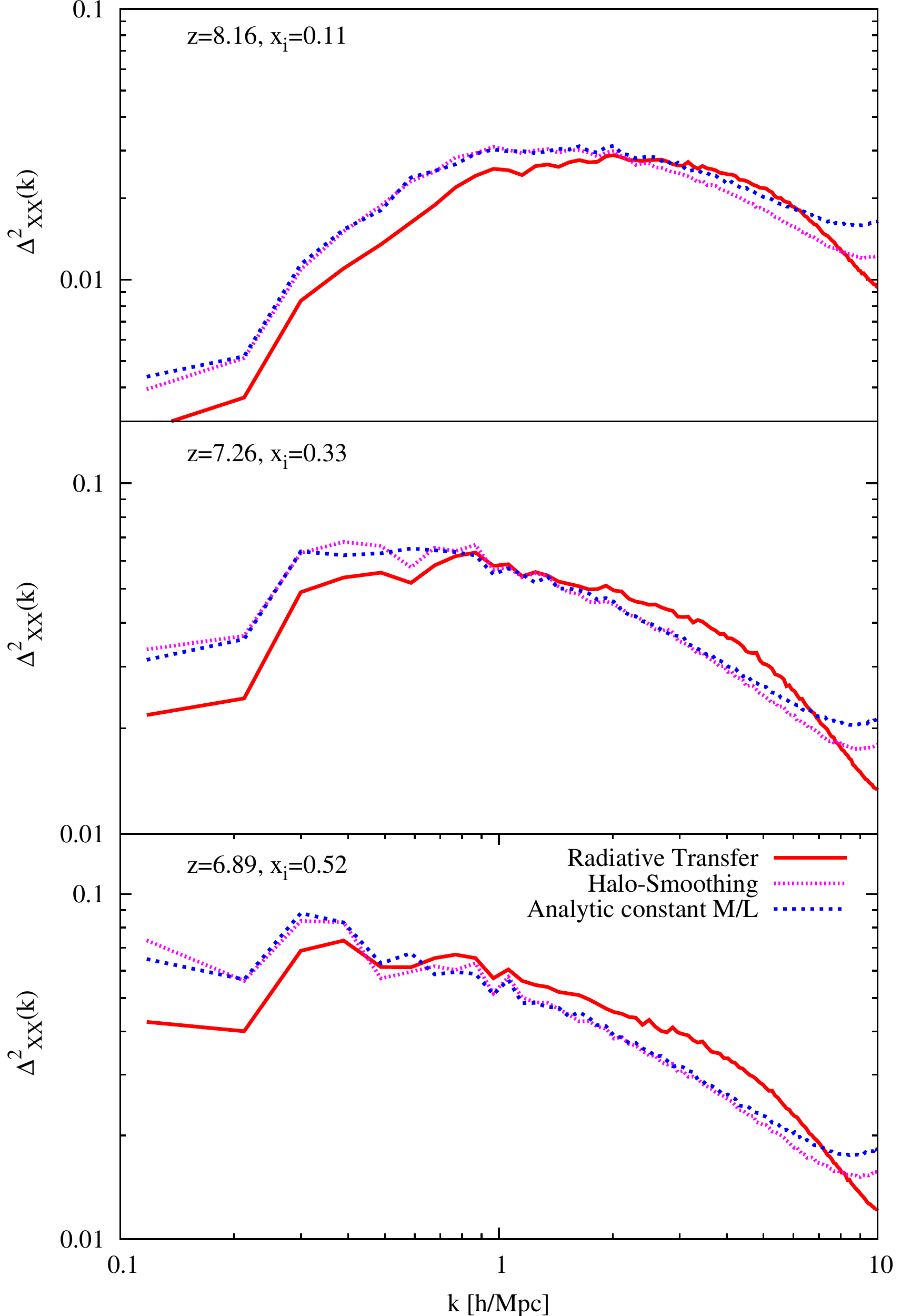}
\caption{Power spectra of the ionized fraction, going from large
  redshift (small ionization fraction) to 
  small redshift (large ionization fraction). The red lines are from
  the radiative transfer simulation, the blue dashed lines are from the
  analytic hybrid calculation, while the purple dotted lines show results from the improved
  scheme of the next section.
   The high-$k$ behavior
  ($k \gtrsim 10 h$ Mpc$^{-1}$) is an artifact from discreteness noise.}
\label{fig:po.xx}
\ec
\end{figure}

For further comparison, we measure the (spherically averaged) 3d ionization power spectrum
as a function of redshift. We consider the ionization field 
$\delta_x = x(\vec{r}) - \avg{x}$, where $x(\vec{r})$ denotes the ionization at
spatial position $\vec{r}$, and $\avg{x}$ denotes the
volume-averaged ionization. Note that we do not normalize by the mean ionization here, 
i.e. we consider the absolute ionization fluctuation, rather than the fractional fluctuation.
The result of the power spectrum calculation is shown in Figure \ref{fig:po.xx}, with power spectra calculated from the radiative transfer simulation
plotted in red. Throughout this paper we plot the dimensionless power spectrum, 
$\Delta^2(k) = k^3 P(k)/(2 \pi^2)$, which yields the contribution to the variance per logarithmic interval
in $k$.
On large scales at high redshift the ionization power
spectrum 
is proportional to the density power spectrum, while it turns over or flattens
on scales in which there are ionized bubbles. On intermediate scales the power spectrum from the radiative transfer simulation has a somewhat larger amplitude. We attribute this to a superior tracking of the density field around the edges of the bubbles: rays can travel into underdense regions and will be hindered by overdensities. In the numeric schemes the features of the density field are washed out somewhat, resulting into a generally smoother edge structure, see also Figure \ref{fig:maps.xx}.
The bubble `feature' moves to progressively
larger scales (small $k$) as reionization proceeds, a further illustration of the
bubble growth seen in Figure \ref{fig:bubble_pdf}. The blue dashed
curves show power spectra from our  
hybrid simulation, which are similar to the radiative transfer power spectra, except with 
slightly more large scale power, and slightly less small scale power. One can also infer
from the figure that an even larger volume simulation is preferable,
in order to better sample the 
large scale ionization power spectrum. Finally, the purple dotted lines are
from an improved numerical scheme which we discuss in the next section.

\begin{figure}
\bc
\includegraphics[width=6.2cm,angle=-90]{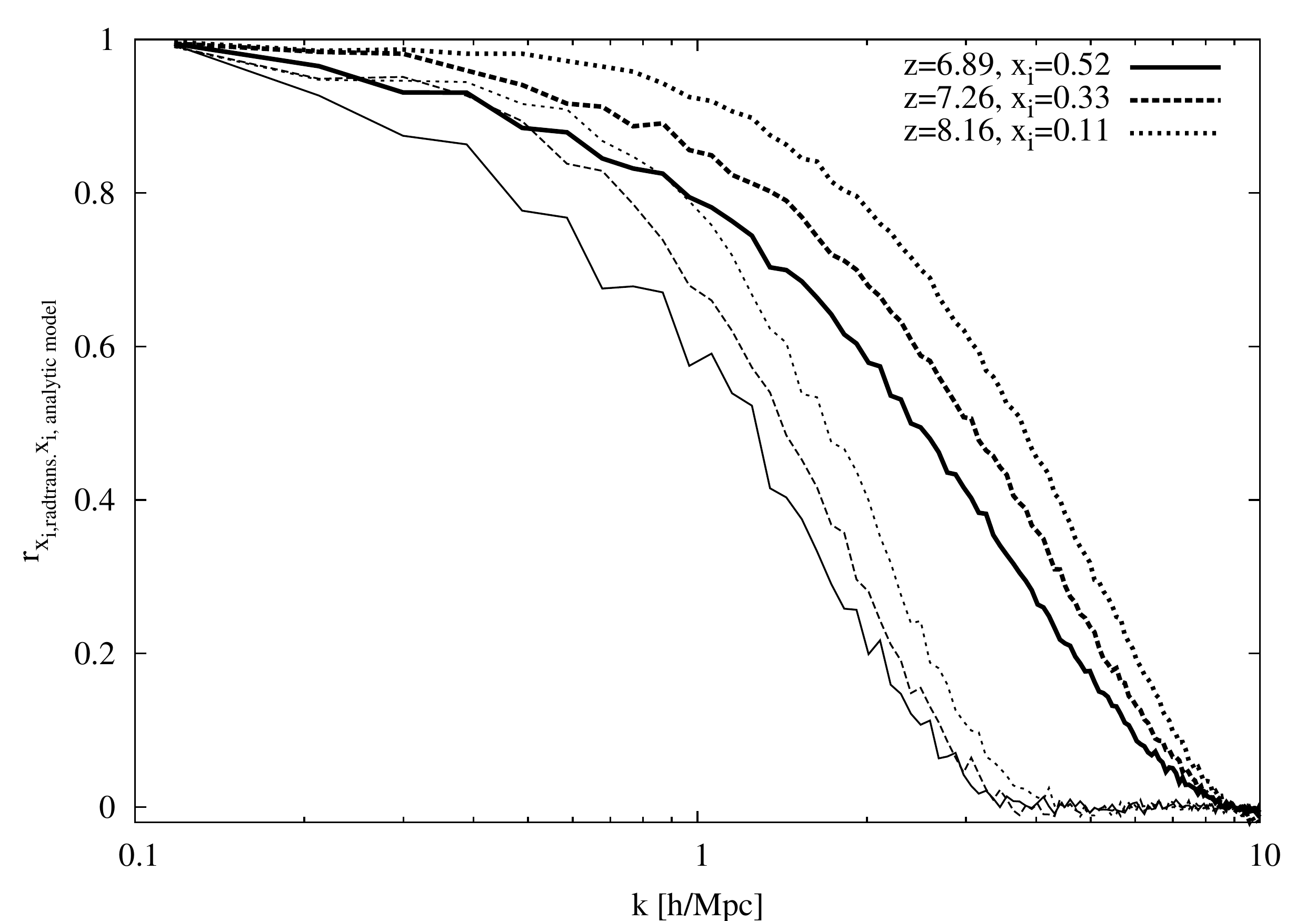}
\caption{Cross correlation coefficient between the ionization fields from the radiative transfer simulation and the analytic model calculations. The thin lines show the cross correlation
  coefficient between the radiative transfer and hybrid simulations at a few different redshifts.
  The thick lines show corresponding results from the improved hybrid simulation described
  in the next section.}
\label{fig:cc.xx}
\ec
\end{figure}

In order to further quantify the agreement between the radiative transfer simulation and the hybrid scheme, we calculate the cross correlation coefficient between the two ionization
fields. The cross
correlation coefficient is defined by
$r(k) = \Delta^2_{\rm x1,x2}(k)/\left[\Delta^2_{\rm x1}(k) \Delta^2_{\rm x2}(k)\right]^{1/2}$.
In this equation $\Delta^2_{\rm x1,x2}(k)$ is the cross power spectrum
between the radiative transfer simulation 
and hybrid scheme ionization fields, while $\Delta^2_{\rm x1}(k)$, and
$\Delta^2_{\rm x2}(k)$ 
are their respective power spectra.  The cross correlation coefficient
is bounded between 
$1$ and $-1$, with $r(k)=1$ indicating perfectly correlated modes, and
$r(k)=-1$ designating 
perfectly anti-correlated modes. The results of this calculation are
shown as thin lines in Figure \ref{fig:cc.xx} (ignore, for now,
the thick lines which show 
results from the improved hybrid scheme introduced in the next
section). The correlation  
coefficient is always larger than
$r \sim 0.5$ for scales larger than $k \lesssim 1 h$ Mpc$^{-1}$, while
it drops off on smaller scales. 
 This quantifies the qualitative agreement 
suggested by Figure \ref{fig:maps.xx}: the radiative transfer and hybrid scheme
ionization fields trace each 
other closely on scales larger than $k \lesssim 1 h$ Mpc$^{-1}$.
The cross correlation between the two fields becomes slightly weaker
at low redshift, as the average ionization increase. A plausible explanation for the
slightly worse agreement at low redshift
is that our hybrid simulation scheme has difficulty with `bubble mergers' (see the Appendix), which
are more frequent at high ionization fraction.

Why does the cross correlation between the two fields drop off around 
$k \gtrsim 1 h$ Mpc$^{-1}$? 
The analytic model assumes a one-to-one correspondence between the
abundance of halos and  
the (Lagrangian) matter overdensity on a given smoothing scale. We
know this is inexact. 
For one, the abundance of our minimum mass sources is $M dn/dM
\lesssim 1$ Mpc$^{-3}$. On 
$\sim 1$ Mpc scales, we therefore expect significant Poisson scatter
in the abundance of  
ionizing sources in our radiative transfer simulation (see also \citealt{Furlanetto:2005ax,Cohn:2006ge}). To explore this further, we compute the cross  
power spectrum between the halo density field and the matter density
field. The cross correlation 
coefficient between the halo and matter density fields qualitatively
mirrors the cross 
correlation between the two ionization fields seen in Figure
\ref{fig:cc.xx}, dropping off 
at $k \gtrsim 1 h$ Mpc$^{-1}$. In other words,
the \emph{halo bias is stochastic} on scales of $k \gtrsim 1 h$
Mpc$^{-1}$ for our assumed 
source population. 
This stochasticity is not incorporated in our analytic hybrid scheme, and 
likely leads to the lack of small scale structure compared to the
ionization field simulated through radiative transfer.      
We will return to this issue in \S \ref{sec:improved_mc}. We note
here, however, that this Poisson 
scatter would presumably be less important if our radiative transfer simulation resolved
smaller, more abundant galaxies. 

\begin{figure}
\bc
\includegraphics[width=9cm]{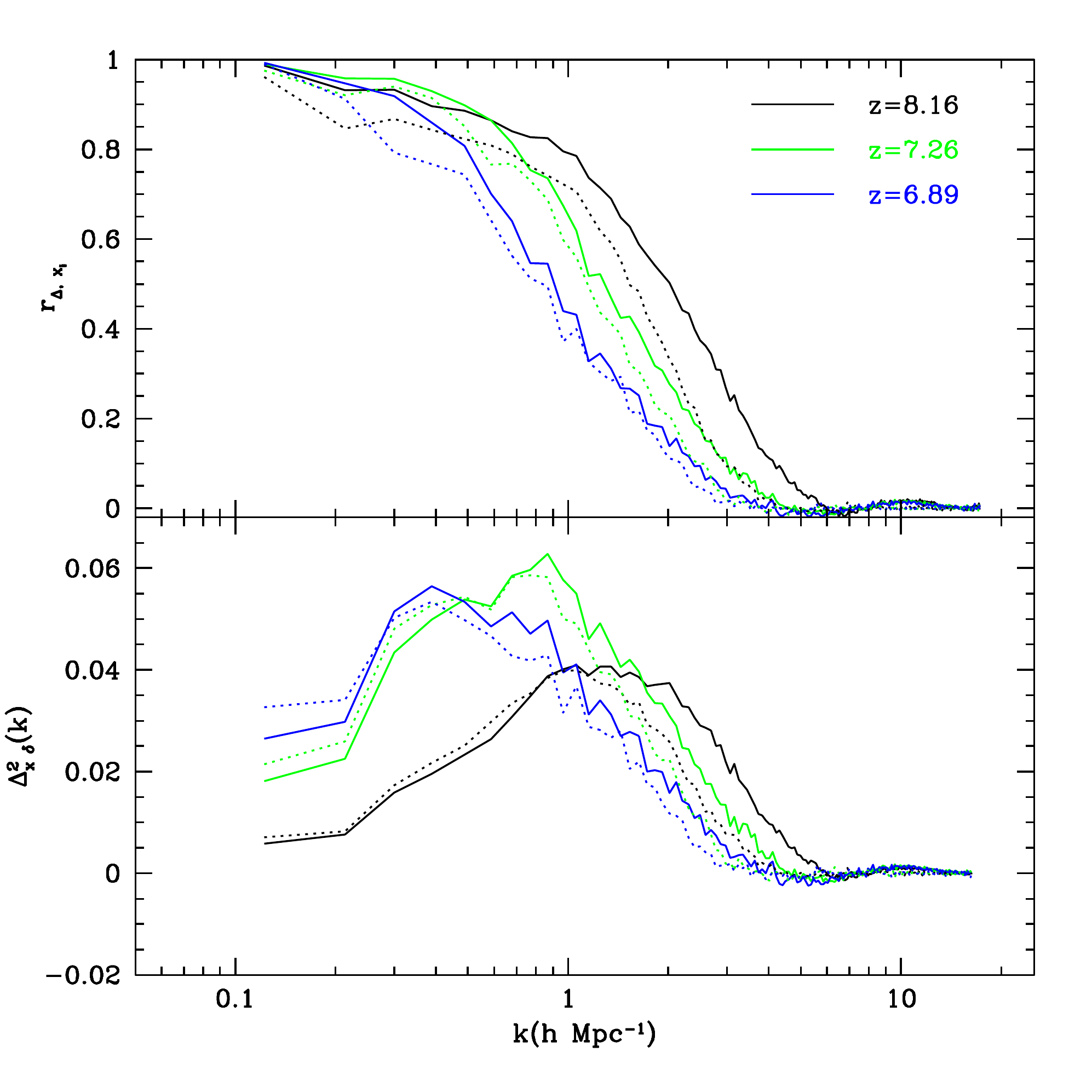}
\caption{\emph{Top panel:} Cross correlation coefficient between
the ionization and density field. The solid (dotted) lines show the cross 
correlation coefficient between the ionization and density fields in the radiative transfer
simulation (hybrid scheme) at several redshifts.
\emph{Bottom panel:} Cross-power spectrum between the ionization and density
field. Solid lines are calculations from the radiative transfer simulation, while dotted lines
are from the hybrid scheme.}
\label{fig:rco_xd}
\ec
\end{figure}

The analytic model connects ionized regions with large scale
overdensities, which contain more sources and are reionized before
underdense regions  
(FZH04, \citealt{Barkana:2003qk}).
The model therefore predicts that the ionization field is positively correlated
with the matter density (e.g. McQuinn et al. 2005b),
before turning over on scales comparable to
that of the ionized bubbles (FZH04). 
Figure \ref{fig:rco_xd} shows the cross power spectrum between
ionization and density (bottom 
panel) as well as the cross-correlation coefficient between the two
fields. The radiative transfer simulation 
results (solid lines) nicely mirror the analytic model predictions (dotted
lines). Since the analytic model ionization field is based on the initial condition density field, it is slightly less correlated with the evolved density field than the radiative transfer simulation ionization field. In our radiative transfer simulation and hybrid
scheme, reionization proceeds inside-out
with the overdense  
regions reionized before underdense regions,
as emphasized by FZH04 and Sokasian et al. (2003, 2004).
Recombinations, underestimated in our present simulations, 
could potentially weaken this correlation or, in an
extreme case, 
reverse the correlation with voids ionized first \citep{Miralda-Escude:2000}. We intend to explore
this in future work. 

\section{Improved numerical scheme}
\label{sec:improved_mc}

Although the agreement between our radiative transfer simulation and the hybrid scheme is already quite good, 
we present here a modified numerical scheme that improves upon the one presented
in \S \ref{sec:analytic} and \citet{Zahn:2005fn}. Specifically, we aim to fix two short-comings
of the analytic calculation. First, as mentioned previously, the analytic
calculation is based on the 
Press-Schechter formula for the collapse fraction. This formula is
derived assuming sharp $k$-space 
filtering, while our scheme filters the initial density field with a top-hat in real space, which
is slightly inconsistent (\citealt{Mcquinn:2005ce}, the Appendix). 
Second, the mass function in our radiative transfer simulation (Figure
\ref{fig:mass_func}) is closer to the
\citet{Sheth:1999mn} mass function than the \citet{Press:1973iz} mass function. Finally, the 
analytic calculation assumes a one-to-one correspondence
between initial over-density and halo abundance. As we discussed in \S
\ref{sec:pspec}, the halo bias 
in our N-body simulation is \emph{stochastic} on small scales.

Each of these shortcomings can be remedied by directly using the
simulated halos in our numerical scheme, rather than the Press-Schechter formula for the collapse
fraction. More specifically, we place the halo distribution from our N-body simulation on a grid
and compare, at each grid cell, the halo mass to the total mass enclosed by a spherical top-hat. We 
then use a condition
analogous to Equation (\ref{eq:barrier}) to determine whether a region is ionized by the sources within it. 
In other words, the calculation proceeds exactly as in \S \ref{sec:analytic}, except that we
use the halo distribution directly from the simulation, rather than Press-Schechter theory.   Note further that 
we now consider the evolved, non-linear density field rather than the initial, linear density field to
determine if a region can self-ionize. We will call this improved numerical implementation the `halo-smoothing' scheme in what follows. The CPU intensity of this scheme is again dominated by the number of FFT's necessary to achieve convergence in the bubble size statistic. As with the analytic scheme, this is roughly 12 minutes on a 3GHz Intel Xeon desktop computer.

The results of this new scheme are shown in comparison with radiative transfer and analytic calculation in Figure \ref{fig:maps.21}, where
we show 21 cm brightness temperature fluctuations (see \S \ref{sec:21cm}) for a thin slice 
through the simulation volume. This is analogous to Figure \ref{fig:maps.xx}, except the ionized
regions are now dark, the neutral regions now bright, and fluctuations in the gas density are
now visible in the neutral regions.  
The \emph{left column}
shows results from our radiative transfer simulation, the \emph{right column} shows the standard
FZH04-type implementation, while the \emph{center column} shows our improved halo-smoothing scheme.  
The blue dots in the
left and center column show the ionizing sources contained in the thin simulation slice. 
The new scheme clearly resembles the full simulation more closely, with more disconnected ionized regions,
owing to the presence of Poisson fluctuations in the source distribution.

Figure \ref{fig:po.xx} quantitatively illustrates improved agreement with the radiative transfer calculation,
with our improved scheme showing more small scale power than the hybrid simulation scheme.
Figure
\ref{fig:cc.xx} additionally shows the cross correlation between the radiative transfer ionization field
and the ionization field in the improved numerical scheme (thick lines).
The halo-smoothing ionization field traces the ionization field simulated through radiative transfer more closely, and down to smaller
scales, than in our initial calculation. We attribute the improved agreement largely to our incorporation,
in the improved scheme, of Poisson scatter in the halo abundance.

\begin{figure*}
\bc
\includegraphics[width=18cm]{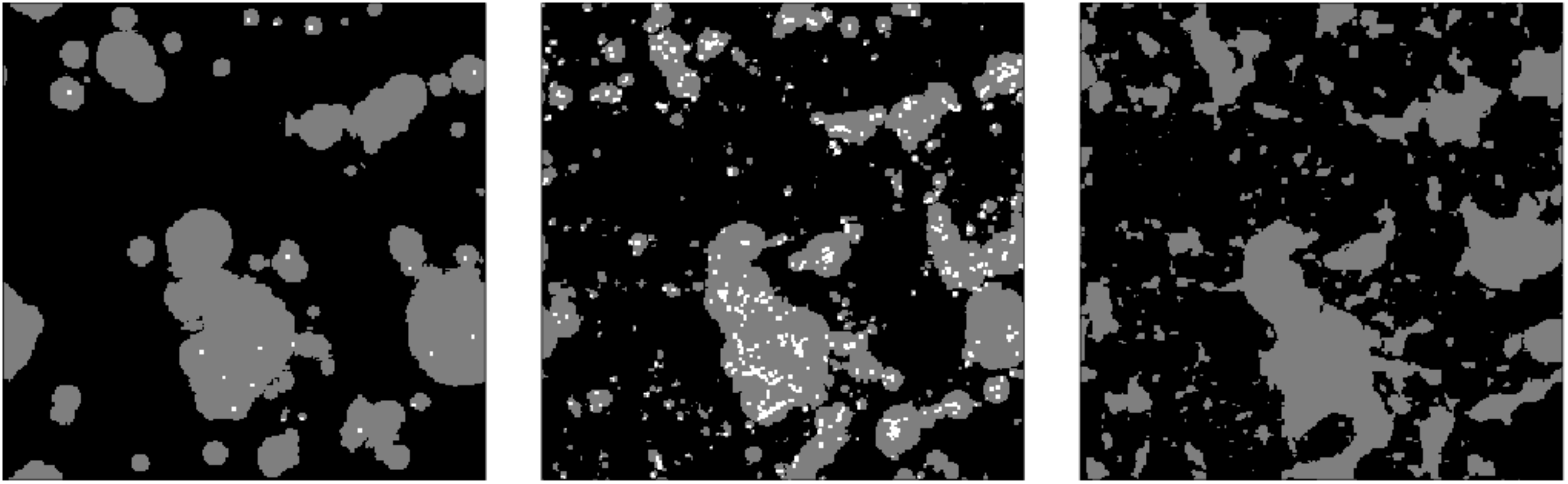}
\caption{Dependence of reionization morphology on source density. In the \emph{left panel} we show the ionization field from our halo-smoothing procedure using only sources (white points) with mass larger than $M \geq 4 \times 10^{10} M_\odot$ (note that some sources contributing to the ionized regions lie in front or behind the thin slice shown). With this choice, the number density of sources roughly matches that of  $M \geq 2 \times 10^9 M_\odot$ sources at $z \sim 14$ (as in \citealt{Iliev:2005sz}). The \emph{center panel} shows the result with our usual source prescription, indicating a significantly more complex morphology. Finally the \emph{right panel} shows, for comparison, the analytic model with $M_{\rm min}=10^8 M_\odot$. Each panel is at $z=7.26$, and in each case the source efficiencies are adjusted to match $x_{\rm i, v} = 0.33$.}
\label{fig:maps.mmin}
\ec
\end{figure*}

If the ionizing sources are even less abundant than we assume presently, the
Poisson scatter naturally becomes more important.
Indeed for sufficiently rare sources, Poisson fluctuations dominate over source
clustering on the scale of a typical bubble, and bubble growth is less `collective' than
in our fiducial model.
In this regime, the morphology of HII regions during reionization may be qualitatively
different.
To examine this, we repeat our halo-smoothing calculation at $z=7.26$ including only
halos with $m \geq 4 \times 10^{10} M_\odot$ as sources. We adjust
the ionizing efficiency of these rarer sources upward to match our usual ionized
fraction at this redshift, $x_{\rm i, v}  = 0.33$,
in order to compare maps at fixed ionization fraction. The result of this
calculation is shown in the left panel of Figure \ref{fig:maps.mmin}. One can
see that the bubbles are considerably more spherical than in our usual source
prescription (middle panel), and that the HII regions have a more sharply defined
scale. The left panel further illustrates that for this source prescription there are very
few sources in each bubble. Note that this is a thin slice, and some sources contributing to bubble growth lie above or below it.

Furthermore, the left panel qualitatively resembles the morphology seen in the
reionization simulations of \citet{Iliev:2005sz} (see their Figure 8\footnote{Note
for comparison with these authors' figure: their simulation box has a side length
of $L = 100$ Mpc/$h$, while ours has $L= 65.6$ Mpc/$h$.}).
Their simulations are done at higher redshift, but have a similar
source number density as our present, extreme choice of $m \geq 4 \times 10^{10} M_\odot$ (with this choice our simulation volume contains roughly 5,000 sources at z=7.26).  We regard the morphology
seen in \citet{Iliev:2005sz} as unlikely to represent the true morphology
of HII regions during reionization. Their choice of minimum source mass 
($M_{\rm min} = 2.5 \times 10^{9} M_\odot$) is driven
by the low mass resolution of their simulations, and the efficiency of their ionizing
sources is boosted extremely high in order to match first-year WMAP constraints \citep{Kogut:2003et}. In other words, their simulation represents a very extreme
case of reionization by rare, bright sources.
Our simulation is also missing plausible ionizing sources,
given our comparable minimum source mass. However, owing to the different assumptions about the ionizing efficiency in our simulation, reionization occurs
later and so our sources are much more abundant (265,000 sources in the simulation volume at z=7.26). We are hence
still in the regime where HII regions grow collectively, and we expect only small
modifications to the morphology and size distribution of HII regions when we include still smaller mass sources.
This is illustrated in the right panel of Figure \ref{fig:maps.mmin} where we show
predictions for our original hybrid scheme (\S \ref{sec:analytic}), with the minimum
source mass extended down to the cooling mass, $M_{\rm min} \sim 10^{8} M_{\rm sun}$. While there are some differences with the results from our usual source
prescription (center panel), the differences are clearly smaller than in comparison to the Poisson-dominated
case (left panel). \footnote{We note the possibility that feedback effects, which have note been included in our simulations, might suppress the formation of the lowest mass sources and lead in extreme cases to a morphology that resembles that seen in \citet{Iliev:2005sz}.} The differences with \citet{Iliev:2005sz} highlight the utility
of our fast numerical schemes for quickly examining many different prescriptions
for the ionizing sources and for understanding the robustness of the results.

\section{21 cm signal and power spectra}
\label{sec:21cm} 

The statistics discussed in \S \ref{sec:stats} are largely diagnostic, aimed at describing the size
distribution of HII regions in the simulation, and characterizing the agreement between the radiative transfer simulation and analytic calculations. In this section we make a more observationally relevant comparison, contrasting
radiative transfer and analytic 21 cm power spectra.

The 21 cm brightness temperature, relative to the CMB, at observed
frequency, $\nu$, and redshift, $z$,  
is (e.g. \citealt{Zaldarriaga:2003du}):
\beqa
\delta T(\nu)
\, & \approx \, & 26 \, (1+\delta_s) x_H \left( \frac{T_S - T_{\rm
CMB}}{T_S} \right) \left( \frac{\Omega_b h^2}{0.022} \right) \nonumber \\
& & \times \left[ \left(\frac{0.15}{\Omega_m h^2} \right) \, \left(
\frac{1+z}{10} \right) \right]^{1/2} {\rm mK} \, .
\label{eq:Tb}
\nonumber
\eeqa
where $\delta_s$ is the density contrast of gas in redshift space, and $T_S$ is the spin temperature 
of neutral hydrogen. At the redshifts we consider presently, the 21 cm excitation temperature is 
likely coupled to the gas
temperature, and much larger than the temperature of the CMB
(e.g. \citealt{Furlanetto:2006tf,Chen:2003gc,Ciardi:2003hg}), $T_S >> T_{\rm CMB}$, implying $\delta T \propto (1 + \delta_s) x_H$. We then model the 21 cm brightness temperature using the simulated density and peculiar velocity fields, in conjunction
with radiative transfer/analytic calculation simulated ionization fields. 
We incorporate here the effect of redshift space distortions, taking into account the simulated peculiar velocity field.
On large scales, linear infall boosts the spherically averaged 21 cm redshift power spectrum relative to its real space
analogue, analogous to the `Kaiser effect' in galaxy surveys \citep{Kaiser:1987qv, Bharadwaj:2004nr,Mcquinn:2005ak, Barkana:2004zy}. By spherically averaging the signal we lose information about the ionizing sources as well as cosmological parameters, as was discussed e.g. in \citet{Barkana:2004zy}. However, the first generation 21cm experiments will be sensitive mainly in the frequency direction and have difficulty measuring the full angular dependence of the signal \citep{Mcquinn:2005ak}.

\begin{figure}
\bc
\includegraphics[width=9cm]{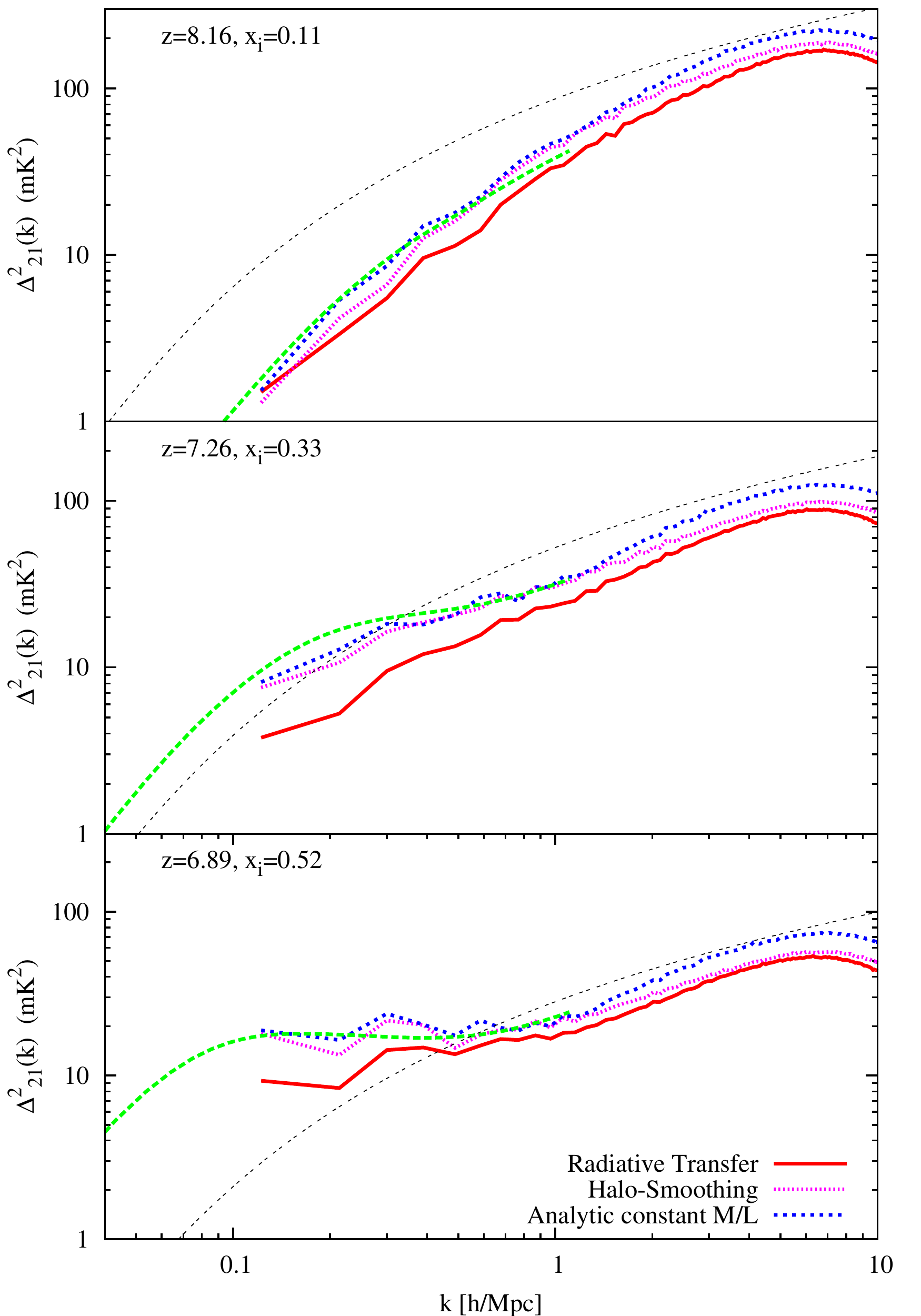}
\caption{The 21 cm brightness temperature power spectra in redshift space.
  The solid red, short-dashed blue and dotted purple lines show the radiative transfer, analytic, and halo-smoothing
  power spectra, respectively. The green long-dashed lines show extrapolations
  of the analytic predictions to large scales. Some of the differences in the predictions
  on large scales may be attributable to our limited simulation volume. The redshift space 21 cm power spectrum approaches $P_\delta x_H^2 1.87$ (shown in the thin dashed curve) on small scales. The differences seen are due to the relevance of higher order contributions to the 21 cm power spectrum (see upcoming work).}
\label{fig:po.21}
\ec
\end{figure}
The result of our power spectrum calculation is shown in Figure \ref{fig:po.21} for three
different redshifts during reionization. 
The results are qualitatively similar to those of Figure \ref{fig:po.xx}, and can be roughly understood
by decomposing the 21 cm power spectrum into three constituent pieces (FZH04):
\beqa
\Delta^2_{\rm 21}(k) &=& \overline{T}_b^2 [\Delta^2_{\rm xx}(k) - \frac{8}{3} \bar{x}_H \Delta^2_{\rm x\delta}(k) +
\frac{28}{15} \bar{x}^2_H \Delta^2_{\rm \delta \delta}(k) ]\,.\nonumber\\
& &
\label{eq:21cm-decomp}
\eeqa
Here $\Delta^2_{\rm xx}$
refers to the ionization power spectrum, 
$\Delta^2_{\rm x \delta}$ refers to the ionization-density cross power spectrum, and 
$\Delta^2_{\rm \delta \delta}$ refers to the density power spectrum. Note that, for illustrative purposes we ignore higher order terms \citep{Mcquinn:2005ak,Furlanetto:2005ax}, although their effects are included in our calculations.  The numerical coefficients in this decomposition come from angle-averaging the redshift space power spectrum.
On scales much larger than the size of the ionized bubbles, each
term in this decomposition is directly proportional to the density power spectrum, and so the 21 cm power spectrum
is directly proportional to the density power spectrum. On the other hand, on very small scales one would expect that the 21 cm power spectrum approaches the density power spectrum multiplied by the neutral fraction squared (and a constant factor $\simeq 1.87$ for the spherically averaged redshift space case). The latter is shown in the thin dashed curves in the Figure. The discrepancy seen is due to the significance of higher order terms that were neglected in Equation \ref{eq:21cm-decomp}, that in reality amount to corrections of order one \citep{Lidz:2006}.

These qualitative trends can be seen in Figure \ref{fig:po.21}. For further illustration, we extrapolate
our predictions to large scales using an analytic model hybrid simulation (green long-dashed lines) which we based on a Gaussian random field with sidelength 300 Mpc/h.

At high redshift, where the ionized regions are small, the 21 cm power spectrum has the shape of the density
power spectrum. At lower redshifts, it begins to flatten on large scales owing to the presence of ionized regions,
before following the shape of the density power spectrum again on small scales. This flattening moves to progressively larger
scales as reionization proceeds, and the bubbles grow larger. 
Our first observational handle on the characteristic sizes of HII regions
at different stages of reionization will likely come from measuring the
21 cm power spectrum, and observing this flattening.
In other work, we will explore the extent 
to which the size distribution of HII
regions can be extracted from future measurements of the 21 cm power spectrum \citep{Zahn:2006fi}.

Notice that the agreement between the analytic and radiative transfer 21 cm power spectra is even better than the agreement  
between the ionization power spectra. While the
ionization field in the radiative transfer simulation has  
more small scale power than the analytic model ionization field, the different approaches show similar amounts of small scale 21 cm power. This owes to the small-scale dominance of the
$\Delta^2_{\rm \delta \delta}(k)$ term in the 21 cm power
spectrum, which overwhelms the difference in small scale ionization power (see Figure \ref{fig:po.xx}).
The 21 cm power spectrum in each analytic scheme seems to provide a very good approximation to the results of our full radiative transfer simulations.
Some of the difference on large scales may be attributable to our limited simulation volume, and a convergence
test with increasing boxsize would be informative, but we leave this to future work.

\begin{figure*}
\bc
\includegraphics[width=18cm]{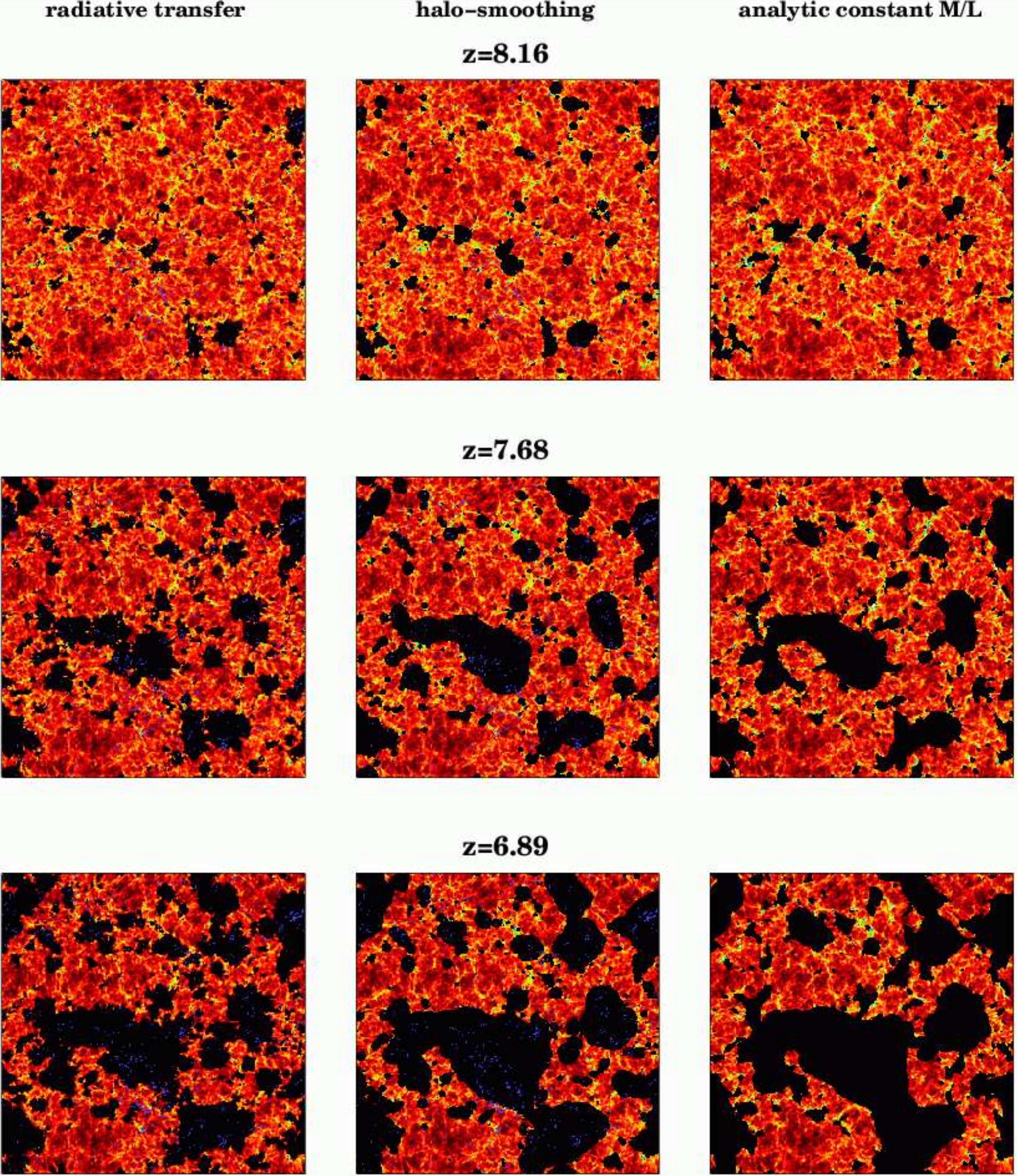}
\caption{21 cm brightness temperature fluctuations. 
  We compare 21 cm maps from the radiative transfer simulation and numerical scheme at  three different redshifts. Each map is $65.6$ Mpc/$h$ on a side, and 
  $0.25$ Mpc/$h$ deep, comparable to the frequency resolution of planned experiments, and shows a different cut then Figure \ref{fig:maps.xx}. The ionized fractions are $x_{\rm i, V}=0.11, 0.33$ and $0.52$ for $z=8.16, 7.26$ and $6.89$ respectively.
 \emph{Left column:}
  Radiative transfer calculation with ionizing
  sources (blue dots). \emph{Middle column:} Halo-smoothing
  procedure (see text) with sources/halos from the N-body simulation. \emph{Right
  column:} Constant mass-to-light ratio version of FZH04, based purely
  on the initial, linear dark matter overdensity.}
\label{fig:maps.21}
\ec
\end{figure*}

\section{Conclusions}
\label{sec:conclusion}

We have presented results from a large volume radiative transfer simulation
and fast numerical schemes based on analytic considerations, and given a  
detailed comparison. Our basic conclusion
is that the approximate schemes agree remarkably well with the radiative transfer simulation. 

Future work should investigate the effect of recombinations which,
we anticipate, will lead to two primary modifications \citep{Furlanetto:2005xx}. First,  
recombinations will slow down reionization by requiring more ionizing
photons to achieve a given ionization fraction.
This should mainly act to modify the redshift evolution of the
ionization fraction, and not 
the size distribution of HII regions at a given ionization fraction,
our main focus in the present 
work. Second, ionization fronts
may be halted upon impacting dense clumps, where the recombination
rate is very high  
(e.g. \citealt{Miralda-Escude:2000,Shapiro:2003gx,Furlanetto:2005xx}). This latter 
effect might, indeed, modify the size distribution of HII regions at a
given ionization fraction. However, 
as long as mini-halos are destroyed by pre-heating prior to
reionization (e.g. \citealt{Oh:2003}), estimates show this 
effect is important only at the tail end of reionization, 
when $x_{\rm i, v} \gtrsim 0.77$ \citep{Furlanetto:2005xx}, which we do not
presently simulate. 

In the future we will address these issues explicitly, along with
other refinements to 
our radiative transfer simulations. 
We intend to consider a more 
sophisticated prescription for the ionizing sources 
\citep{Springel:2002ux,Sokasian:2003au}, and extend the mass range of our sources down to the cooling mass. It will be
interesting to examine how sensitive 
the 21 cm predictions are to the assumed properties of the ionizing
sources \citep{Furlanetto:2005ax}. In particular, in \S \ref{sec:improved_mc}
we found that the morphology and size distribution of HII regions differs
dramatically from our fiducial model when extremely rare, bright sources dominate. This warrants further quantitative investigation. 
Finally, we intend to examine the effect of feedback on reionization,
incorporating Jeans mass suppression 
(e.g. \citealt{Barkana:2000ex,Babich:2005jj,Kramer:2006dn}) in reionized regions of the IGM \citep{Mcquinn:2006rt}.

In spite of these refinements, we contend that the agreement
demonstrated in this paper illustrates 
that the analytic models are on the right track, and provide a useful
complementary tool to radiative 
transfer simulations. The approximate schemes described here are very
fast, allowing quick coverage of 
a large parameter space, convenient for forecasting constraints from
upcoming 21 cm surveys \citep{Zahn:2006fi}. Even full radiative transfer simulations currently have
a large number of free parameters related to the efficiency of the ionizing sources, the
escape fraction of ionizing photons, and sub-grid clumping. Our numerical schemes
allow one to gauge how the expected signal depends on these numerous, unconstrained parameters. 
It can also be used to investigate
non-Gaussianities in the 21 cm signal, as advocated by
\citet{Furlanetto:2004ha},
and 
to construct mock 21 cm survey volumes, providing a useful test of data
analysis procedures, which are presently  
still under development. This is particularly relevant
given that surveys like the MWA 
will be done in large volumes of several co-moving cubic Gigaparsecs,
prohibitive for current radiative transfer
simulations, but manageable with analytic calculations. 
Finally, it
might be interesting to couple 
the fast analytic model schemes with a gas-dynamical calculation to
investigate the impact of reionization 
on galaxy formation.

\section*{Acknowledgments} 

We thank Aaron Sokasian for providing his radiative transfer code, and Volker Springel for providing an enhanced version of Gadget. We also thank Tom Abel, Marcelo Alvarez, Katrin Heitmann, Salman Habib,
and Miguel Morales for helpful conversations. 
The authors are supported by the David and Lucile Packard
Foundation, the Alfred P. Sloan Foundation, and NASA grants
AST-0506556 and NNG05GJ40G. 

\begin{appendix}

\section{Photon Conservation in our approximate simulation schemes}

The objective of this Appendix is to show that the pure FZH04 model conserves photons, but
that our numerical schemes do not precisely conserve photons. We then discuss the implications
of this finding. In the pure FZH04 model, we can prove that the global ionization fraction is 
given by $\bar{x} = \zeta \times f_{\rm coll.}$. This is just a reflection of photon
conservation: as we sum up the total ionized mass from individual HII regions, no photons
are lost or gained in our accounting of the net ionized mass.

A rigorous proof proceeds as follows. For simplicity, we outline this proof using the 
pure FZH04 barrier,
but the proof can be easily generalized to the barrier of Equation (\ref{eq:barrier}). Let
us consider random walks in the $(\delta, \sigma^2)$ plane (e.g. Bond et al. 1991), generated
using top-hat smoothing in $k$-space. 
We consider the
first up-crossing distributions for two types of barriers. First, we examine the 
probability that a random walk crosses the `bubble barrier', representing the critical density 
threshold for a region to self-ionize (see Figure 1 of FZH04).
We denote the differential probability that a random walk crosses this barrier, at a resolution
between $\sigma^2$ and $\sigma^2 + d \sigma^2$, by $dP_b/d\sigma^2$. Next, we consider the
ordinary Press-Schechter barrier, representing the critical overdensity for a region to collapse and
form a halo. The 
differential probability distribution for a random walk to cross the `collapse barrier', at a 
resolution between $\sigma^{\prime 2}$
and $\sigma^{\prime 2} + d \sigma^{\prime 2}$, is denoted by $dP_c/d\sigma^{\prime 2}$. 
Similarly, the probability
distribution for collapse in a region with large-scale overdensity $\delta_b$, on smoothing scale
$\sigma^2$, is denoted by $dP_c(\sigma^{\prime 2} | \delta_b, \sigma^2)/d\sigma^{\prime 2}$.
The total ionized mass in a region of large scale overdensity $\delta_b$, at a smoothing
scale $\sigma^2$, is then given by 
\beq
\int \, dM_h \, \zeta \, M_h \, \frac{d\sigma^{\prime 2}}{dM_h} 
\frac{dP_c(\sigma^{\prime 2} | \delta_b, \sigma^2)}{d\sigma^{\prime 2}}\,.
\label{eq:mass_local}
\eeq
Note that the conditional probability distribution in this formula is calculated by considering
the fraction of random walks, originating at $(\delta_b, \sigma^2)$, that cross the collapse
barrier at higher resolution (Lacey \& Cole 1993). The mass calculated using 
Equation (\ref{eq:mass_local}) is precisely the ionized mass
in an HII region that crosses the `bubble barrier' at the point $(\delta_b, \sigma^2)$.
In order to find the total ionized mass in all HII regions, we merely need to integrate
over all such crossings, i.e., we integrate Equation (\ref{eq:mass_local}) over $\sigma^2$ weighted
by the probability of crossing the bubble barrier. Symbolically, the total ionized mass in
the IGM is then given by 
\beq
\int d \, \sigma^2 \, \frac{dP_b(\sigma^2)}{d\sigma^2} \int dM_h \, \zeta \, M_h \, \frac{d\sigma^{\prime 2}}{dM_h} 
\frac{dP_c(\sigma^{\prime 2} | \delta_b, \sigma^2)}{d\sigma^{\prime 2}}\, .
\eeq
 
This is one expression for the total ionized mass in the IGM, obtained by summing the ionized
mass in all individual HII regions. Our proof of photon conservation is completed by showing that 
this `local' expression
matches a separate expression, proportional to the global collapse fraction. The
total mass in halos is simply 
\beq
\int dM_h \,  M_h \, \frac{d\sigma^{\prime 2}}{dM_h} \frac{dP_c(\sigma^{\prime 2})}{d\sigma^{\prime 2}}\, ,
\eeq
and the total, photon-conserving, ionized mass is just $\zeta$ times this expression. 
Now, this expression, proportional to the global collapse fraction follows
by considering the crossing distribution of the collapse barrier, \emph{irrespective} 
of when each random walk crosses the bubble barrier. This result clearly must match that of
Equation (\ref{eq:mass_local})
since for two random variables, $x$ and $y$ with probability distributions $P(x)$ and $P(y)$, 
$\int dy P(y) \int dx x P(x|y) = \int dx x P(x)$,
i.e. in one case we are integrating (`marginalizing') over `bubble crossings', and in the other case
we are not. This proves that the pure FZH04 model conserves photons, and 
our numerical implementation of the FZH04 model with a sharp $k$-space filter
indeed conserves photons.
 
In practice, however the hybrid scheme of \S \ref{sec:analytic} smoothes the density field with a 
top-hat in real space, rather than a sharp $k$-space filter. In this case photon conservation is
not guaranteed. Specifically, the expression in Equation (\ref{eq:fcoll}) of \S \ref{sec:analytic} is 
rigorously equal to the collapse fraction only
for sharp $k$-space filtering, and not for real-space smoothing (see also \citealt{Mcquinn:2005ce}). One 
option would be to simply apply our
algorithm with a sharp $k$-space filter, but we find that this produces artificial features
in our ionization maps (ringing in configuration space). For this reason, we prefer to apply
our algorithm using a top-hat in real space. In practice this leads to photon non-conservation 
at the $20\%$ level, with our algorithm systematically under-shooting the expected ionization,
$\bar{x} = \zeta \times f_{\rm coll.}$.
To compare with the radiative transfer simulation, we simply boost the ionizing efficiency
to make up for this photon loss, matching the (volume-weighted) ionization fraction in the 
radiative transfer simulation.

\begin{figure}
\bc
\includegraphics[width=15cm]{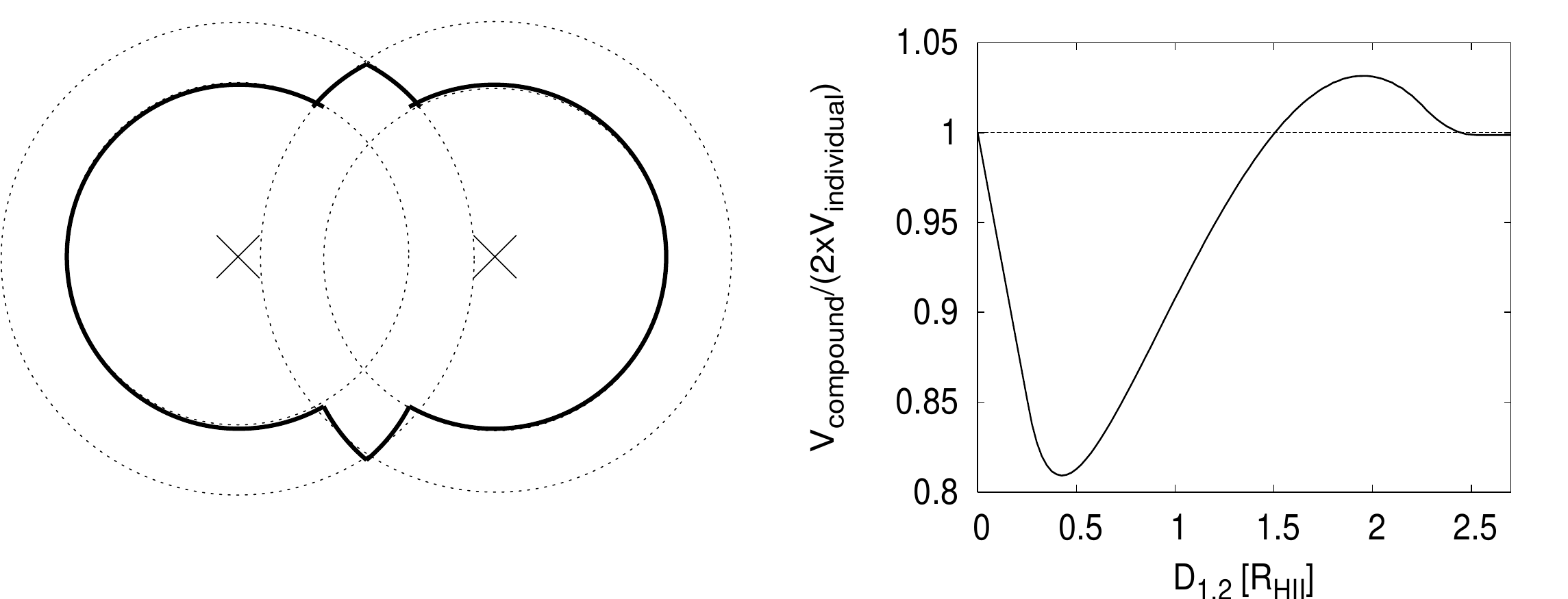}
\caption{An illustration of photon non-conservation in our `halo-smoothing' simulations.
The \emph{left panel} shows the ionized region from our numerical scheme
for the toy problem of two equal-luminosity sources. Our scheme leads to the HII region
denoted by the thick solid boundary. In reality, one expects an oblong HII region, as photons
from each source stream to the edge of the HII region created by the other source, and expand
its volume. For this toy problem, our procedure initially over-estimates the ionized volume by a few percent at moderate source separations, then under-estimates the ionized volume at smaller
 separations. In the limit of non-overlapping HII regions (very large source separations), and in the limit of very small source separations, we recover the expected ionized volume. This is illustrated in the \emph{right panel} which shows the fractional photon loss/gain as a function of source separation. The x-axis is the source separation in units of the radius of an individual HII region.}
 \label{fig:npc}
 \ec
 \end{figure}

Is photon-conservation fulfilled in our improved `halo-smoothing' scheme? We consider a simple toy
problem to illustrate that our improved scheme also does not quite conserve photons. Imagine
two equal luminosity sources in a uniform density field.
When the ionized regions surrounding these sources begin to overlap, the spherical top-hat criterion can lead to somewhat unphysical features. This is sketched in the left panel of Figure \ref{fig:npc}. Our algorithm does not allow for flux from one source to expand the HII region surrounding the second source. Instead of both HII spheres (with initial radius
$r_1$) growing further during overlap, a new ionized region arises between them, the overlap of two spheres with radius $r_2=2^{1/3} r_1$. In Figure \ref{fig:npc} we plot the ratio of the ionized volume in our scheme, to the expected,
photon-conserving ionized volume. The figure clearly illustrates that our scheme generally loses photons as two bubbles `merge'.
The precise level of photon loss in our `halo-smoothing' scheme will depend on the ionized fraction, the size distribution of the HII regions, the luminosity and bias of the sources interior to merging bubbles, and the rate of merging
bubbles. In practice, the level of photon non-conservation in our halo-smoothing scheme is also at the $20\%$ level. 
Again our solution is to uniformly boost the ionizing efficiency of our sources to match the (volume-weighted) ionization fraction in the radiative transfer simulation. Ideally, we would only boost the efficiency in recently merged bubbles where we expect photon
loss. In practice, any error associated with this approximation appears small, although the higher 
frequency of bubble
mergers at late stages of reionization makes our scheme slightly less reliable in this 
regime (see Figure \ref{fig:cc.xx}).

\end{appendix}




\begin{thebibliography}{62}
\expandafter\ifx\csname natexlab\endcsname\relax\def\natexlab#1{#1}\fi

\bibitem[{Abel {et~al.}(1998)Abel, Norman, \& Madau}]{Abel:1998qq}
Abel, T., Norman, M.~L., \& Madau, P. 1998, ApJ, 523, 66,
  \eprint{astro-ph/9812151}

\bibitem[{Abel \& Wandelt(2002)}]{Abel:2001qs}
Abel, T., \& Wandelt, B.~D. 2002, MNRAS, 330, L53, \eprint{astro-ph/0111033}

\bibitem[{Babich \& Loeb(2005)}]{Babich:2005jj}
Babich, D., \& Loeb, A. 2005, \eprint{astro-ph/0509784}

\bibitem[{Barkana \& Loeb(2000)}]{Barkana:2000ex}
Barkana, R., \& Loeb, A. 2000, ApJ, 539, 20, \eprint{astro-ph/0001326}

\bibitem[{Barkana \& Loeb(2002)}]{Barkana:2002sp}
Barkana, R., \& Loeb, A. 2002, ApJ, 578, 1, \eprint{astro-ph/0204139}

\bibitem[{Barkana \& Loeb(2004{\natexlab{a}})}]{Barkana:2003ja}
Barkana, R., \& Loeb, A. 2004{\natexlab{a}}, ApJ, 601, 64, \eprint{astro-ph/0305470}

\bibitem[{Barkana \& Loeb(2004{\natexlab{b}})}]{Barkana:2003qk}
Barkana, R., \& Loeb, A. 2004{\natexlab{b}}, ApJ, 609, 474, \eprint{astro-ph/0310338}

\bibitem[{Barkana \& Loeb(2005)}]{Barkana:2004zy}
Barkana, R., \& Loeb, A. 2005, ApJ, 624, L65, \eprint{astro-ph/0409572}

\bibitem[{Becker {et~al.}(2005)Becker, Sargent, Rauch, \&
  Simcoe}]{Becker:2005vg}
Becker, G.~D., Sargent, W. L.~W., Rauch, M., \& Simcoe, R.~A. 2005, ApJ, 640, 69, \eprint{astro-ph/0511541}

\bibitem[{Bharadwaj \& Ali(2004)}]{Bharadwaj:2004nr}
Bharadwaj, S., \& Ali, S.~S. 2004, MNRAS, 352, 142, \eprint{astro-ph/0401206}

\bibitem[{Bond {et~al.}(1991)Bond, Cole, Efstathiou, \& Kaiser}]{Bond:1990iw}
Bond, J.~R., Cole, S., Efstathiou, G., \& Kaiser, N. 1991, ApJ, 379, 440

\bibitem[{Bowman {et~al.}(2006)Bowman, Morales, \& Hewitt}]{Bowman:2005cr}
Bowman, J.~D., Morales, M.~F., \& Hewitt, J.~N. 2006, ApJ, 638, 20,
  \eprint{astro-ph/0507357}

\bibitem[{Chen \& Miralda-Escude(2004)}]{Chen:2003gc}
Chen, X.-L., \& Miralda-Escude, J. 2004, ApJ, 602, 1, \eprint{astro-ph/0303395}

\bibitem[{Ciardi {et~al.}(2003)Ciardi, Ferrara, \& White}]{Ciardi:2003ia}
Ciardi, B., Ferrara, A., \& White, S. D.~M. 2003, MNRAS, 344, L7,
  \eprint{astro-ph/0302451}

\bibitem[{Ciardi \& Madau(2003)}]{Ciardi:2003hg}
Ciardi, B., \& Madau, P. 2003, ApJ, 596, 1, \eprint{astro-ph/0303249}

\bibitem[{Cohn \& Chang(2006)}]{Cohn:2006ge}
Cohn, J.~D., \& Chang, T.-C. 2006, \eprint{astro-ph/0603438}

\bibitem[{Davis {et~al.}(1985)Davis, Efstathiou, Frenk, \&
  White}]{Davis:1985rj}
Davis, M., Efstathiou, G., Frenk, C.~S., \& White, S. D.~M. 1985, ApJ, 292, 371

\bibitem[{{Eisenstein} \& {Hu}(1999)}]{Eisenstein:1997jh}
{Eisenstein}, D.~J., \& {Hu}, W. 1999, \apj, 511, 5,
  \adsurl{http://adsabs.harvard.edu/cgi-bin/nph-bib_query?bibcode=1999ApJ...51
1....5E&db_key=AST}, \eprint{astro-ph/9710252}

\bibitem[{Fan {et~al.}(2005)}]{Fan:2005es}
Fan, X.-H., {et~al.} 2005, \eprint{astro-ph/0512082}

\bibitem[{Furlanetto {et~al.}(2004{\natexlab{a}})Furlanetto, Zaldarriaga, \&
  Hernquist}]{Furlanetto:2004nh}
Furlanetto, S., Zaldarriaga, M., \& Hernquist, L. 2004{\natexlab{a}}, ApJ, 613, 1, \eprint{astro-ph/0403697} [FZH04]

\bibitem[{Furlanetto {et~al.}(2004{\natexlab{b}})Furlanetto, Zaldarriaga, \&
  Hernquist}]{Furlanetto:2004ha}
Furlanetto, S., Zaldarriaga, M., \& Hernquist, L. 2004{\natexlab{b}}, ApJ, 613, 16, \eprint{astro-ph/0404112}

\bibitem[{Furlanetto {et~al.}(2006{\natexlab{a}})Furlanetto, McQuinn, \&
  Hernquist}]{Furlanetto:2005ax}
Furlanetto, S.~R., McQuinn, M., \& Hernquist, L. 2006{\natexlab{a}}, MNRAS,
  365, 115, \eprint{astro-ph/0507524}

\bibitem[{Furlanetto \& Oh(2005)}]{Furlanetto:2005xx}
Furlanetto, S.~R., \& Oh, S.~P. 2005, MNRAS, 363, 103, \eprint{astro-ph/0505065}

\bibitem[{Furlanetto {et~al.}(2006{\natexlab{b}})Furlanetto, Zaldarriaga, \&
  Hernquist}]{Furlanetto:2005ir}
Furlanetto, S.~R., Zaldarriaga, M., \& Hernquist, L. 2006{\natexlab{b}}, MNRAS,
  365, 1012, \eprint{astro-ph/0507266}

\bibitem[{Furlanetto (2006)}]{Furlanetto:2006tf}
  S.~Furlanetto, \eprint{astro-ph/0604040}

\bibitem[{Gnedin(2000)}]{Gnedin:2000uj}
Gnedin, N.~Y. 2000, ApJ, 542, 535, \eprint{astro-ph/0002151}

\bibitem[{Haiman {et~al.}(2000)Haiman, Abel, \& Madau}]{Haiman:2000pd}
Haiman, Z., Abel, T., \& Madau, P. 2000, ApJ, 551, 599, \eprint{astro-ph/0009125}

\bibitem[{Haiman {et~al.}(1997)Haiman, Rees, \& Loeb}]{Haiman:1996rc}
Haiman, Z., Rees, M.~J., \& Loeb, A. 1997, ApJ, 476, 458,
  \eprint{astro-ph/9608130}

\bibitem[{Heitmann {et~al.}(2006)Heitmann, Lukic, Habib, \&
  Ricker}]{Heitmann:2006eu}
Heitmann, K., Lukic, Z., Habib, S., \& Ricker, P.~M. 2006,
  \eprint{astro-ph/0601233}

\bibitem[{Iliev {et~al.}(2005)}]{Iliev:2005sz}
Iliev, I.~T., {et~al.} 2005, \eprint{astro-ph/0512187}

\bibitem[{Kaiser(1987)}]{Kaiser:1987qv}
Kaiser, N. 1987, MNRAS, 227, 1

\bibitem[{Kneib {et~al.}(2004)Kneib, Ellis, Santos, \& Richard}]{Kneib:2004dq}
Kneib, J.-P., Ellis, R.~S., Santos, M.~R., \& Richard, J. 2004, ApJ, 607, 697,
  \eprint{astro-ph/0402319}

\bibitem[{Kogut {et~al.}(2003)}]{Kogut:2003et}
Kogut, A., {et~al.} 2003, ApJS, 148, 161, \eprint{astro-ph/0302213}

\bibitem[{Kohler {et~al.}(2005)Kohler, Gnedin, \& Hamilton}]{Kohler:2005gg}
Kohler, K., Gnedin, N.~Y., \& Hamilton, A. J.~S. 2005,
  \eprint{astro-ph/0511627}

\bibitem[{Kramer {et~al.}(2006)}]{Kramer:2006dn}
  Kramer, R.~H., Haiman, Z. and Oh, S.P. 2006,
  \eprint{astro-ph/0604218}
  
\bibitem[{{Lacey} \& {Cole}(1993)}]{Lacey:1993}
{Lacey}, C., \& {Cole}, S. 1993, \mnras, 262, 627,
  \adsurl{http://adsabs.harvard.edu/cgi-bin/nph-bib_query?bibcode=1993MNRAS.262..627L&db_key=AST}

\bibitem[{Lewis(2006)}]{Lewis:2006ma}
Lewis, A. 2006, \eprint{astro-ph/0603753}

\bibitem[{Lidz {et. al}(2006)}]{Lidz:2006}
Lidz, A. Zahn, O., Zaldarriaga, M., McQuinn, M. 2006, in preparation

\bibitem[{Loeb {et~al.}(2005)Loeb, Barkana, \& Hernquist}]{Loeb:2004zs}
Loeb, A., Barkana, R., \& Hernquist, L. 2005, ApJ, 620, 553,
  \eprint{astro-ph/0403193}

\bibitem[{Madau {et~al.}(1996)Madau, Meiksin, \& Rees}]{Madau:1996cs}
Madau, P., Meiksin, A., \& Rees, M.~J. 1996, ApJ, 474, 429, \eprint{astro-ph/9608010}

\bibitem[{Malhotra \& Rhoads(2005)}]{Malhotra:2005qf}
Malhotra, S., \& Rhoads, J. 2005, \eprint{astro-ph/0511196}

\bibitem[{McQuinn {et~al.}(2005)McQuinn, Furlanetto, Hernquist, Zahn, \&
  Zaldarriaga}]{Mcquinn:2005ce}
McQuinn, M., Furlanetto, S.~R., Hernquist, L., Zahn, O., \& Zaldarriaga, M.
  2005, ApJ, 630, 643, \eprint{astro-ph/0504189}

\bibitem[{McQuinn {et~al.}(2005)McQuinn, Zahn, Hernquist, Zaldarriaga, \&
  Furlanetto}]{Mcquinn:2005ak}
McQuinn, M., Zahn, O., Hernquist, L., Zaldarriaga, M., \& Furlanetto, S. R.,
  \eprint{astro-ph/0512263}

\bibitem[{McQuinn {et~al.}(2006) in prep.}]{Mcquinn:2006rt}
McQuinn, M., et. al, in preparation
 
\bibitem[{Mesinger {et~al.} (2004)}]{Mesinger:2004rk}
  A.~Mesinger and Z.~Haiman,
   ``Evidence for a Boundary of the Cosmological Stromgren Sphere and for
  arXiv:astro-ph/0406188.
  
\bibitem[{Mellema {et~al.}(2005)Mellema, Iliev, Alvarez, \&
  Shapiro}]{Mellema:2005ht}
Mellema, G., Iliev, I., Alvarez, M., \& Shapiro, P. 2005,
  \eprint{astro-ph/0508416}

\bibitem[{Mellema {et~al.}(2006)Mellema, Iliev, Pen, \&
  Shapiro}]{Mellema:2006pd}
Mellema, G., Iliev, I.~T., Pen, U.-L., \& Shapiro, P.~R. 2006,
  \eprint{astro-ph/0603518}

\bibitem[{{Miralda-Escud{\'e}} {et~al.}(2000){Miralda-Escud{\'e}}, {Haehnelt},
  \& {Rees}}]{Miralda-Escude:2000}
{Miralda-Escud{\'e}}, J., {Haehnelt}, M., \& {Rees}, M.~J. 2000, \apj, 530, 1,
  \adsurl{http://adsabs.harvard.edu/cgi-bin/nph-bib_query?bibcode=2000ApJ...53
0....1M&db_key=AST}

\bibitem[{Oh(2002)}]{Oh:2002ry}
Oh, S.~P. 2002, MNRAS, 336, 1021, \eprint{astro-ph/0201517}

\bibitem[{Oh \& Haiman(2003)}]{Oh:2003}
Oh, S.~P., \& Haiman, Z. 2003, MNRAS, 346, 456

\bibitem[{Page {et~al.}(2006)}]{Page:2006hz}
Page, L., {et~al.} 2006, \eprint{astro-ph/0603450}

\bibitem[{Pen {et~al.}(2004)Pen, Wu, \& Peterson}]{Pen:2004de}
Pen, U.-L., Wu, X.-P., \& Peterson, J. 2004, AAS, 205, 131, \eprint{astro-ph/0404083}

\bibitem[{Press \& Schechter(1974)}]{Press:1973iz}
Press, W.~H., \& Schechter, P. 1974, ApJ, 187, 425

\bibitem[{{Razoumov} {et~al.}(2002){Razoumov}, {Norman}, {Abel}, \&
  {Scott}}]{Razoumov:2002}
{Razoumov}, A.~O., {Norman}, M.~L., {Abel}, T., \& {Scott}, D. 2002, \apj, 572,
  695,
  \adsurl{http://adsabs.harvard.edu/cgi-bin/nph-bib_query?bibcode=2002ApJ...57
2..695R&db_key=AST}

\bibitem[{Reed {et~al.}(2005)}]{Reed:2003hp}
Reed, D., {et~al.} 2005, MNRAS, 357, 82, \eprint{astro-ph/0312544}

\bibitem[{Rhoads {et~al.}(2003)}]{Rhoads:2002dh}
Rhoads, J.~E., {et~al.} 2003, AJ, 125, 1006, \eprint{astro-ph/0209544}

\bibitem[{Santos {et~al.}(2003)Santos, Cooray, Haiman, Knox, \&
  Ma}]{Santos:2003jb}
Santos, M.~G., Cooray, A., Haiman, Z., Knox, L., \& Ma, C.-P. 2003,
ApJ, 598, 756

\bibitem[{Seljak {et al.}(2006)}]{Seljak:2006bg}
  U.~Seljak, A.~Slosar and P.~McDonald,
  arXiv:astro-ph/0604335.
  
\bibitem[{Shapiro {et~al.}(2004)Shapiro, Iliev, \& Raga}]{Shapiro:2003gx}
Shapiro, P.~R., Iliev, I.~T., \& Raga, A.~C. 2004, MNRAS, 348, 753,
  \eprint{astro-ph/0307266}

\bibitem[{Sheth \& Tormen(1999)}]{Sheth:1999mn}
Sheth, R.~K., \& Tormen, G. 1999, MNRAS, 308, 119, \eprint{astro-ph/9901122}

\bibitem[{Sokasian {et~al.}(2003)Sokasian, Abel, Hernquist, \&
  Springel}]{Sokasian:2003au}
Sokasian, A., Abel, T., Hernquist, L., \& Springel, V. 2003, MNRAS, 344, 607,
  \eprint{astro-ph/0303098}

\bibitem[{{Sokasian} {et~al.}(2001){Sokasian}, {Abel}, \&
  {Hernquist}}]{Sokasian:2001na}
{Sokasian}, A., {Abel}, T., \& {Hernquist}, L.~E. 2001, New Astronomy, 6, 359,
  \adsurl{http://adsabs.harvard.edu/cgi-bin/nph-bib_query?bibcode=2001NewA....6..359S&db_key=AST}

\bibitem[{Sokasian {et~al.}(2002)Sokasian, Abel, \&
  Hernquist}]{Sokasian:2001xh}
Sokasian, A., Abel, T., \& Hernquist, L.~E. 2002, MNRAS, 332, 601,
  \eprint{astro-ph/0112297}

\bibitem[{Sokasian {et~al.}(2004)Sokasian, Yoshida, Abel, Hernquist, \&
  Springel}]{Sokasian:2003gf}
Sokasian, A., Yoshida, N., Abel, T., Hernquist, L., \& Springel, V. 2004,
  MNRAS, 350, 47, \eprint{astro-ph/0307451}

\bibitem[{Spergel {et~al.}(2006)}]{Spergel:2006hy}
Spergel, D.~N., {et~al.} 2006, \eprint{astro-ph/0603449}

\bibitem[{Springel \& Hernquist (2002)}]{Springel:2002ux}
  V.~Springel and L.~Hernquist, 2003,
    MNRAS, 339, 312,
  \eprint{arXiv:astro-ph/0206395}
  
\bibitem[{Springel(2005)}]{Springel:2005mi}
Springel, V. 2005, MNRAS, 364, 1105, \eprint{astro-ph/0505010}

\bibitem[{Totani {et~al.}(2005)}]{Totani:2005ng}
Totani, T., {et~al.} 2005, \eprint{astro-ph/0512154}

\bibitem[{Wyithe {et~al.}(2004)}]{Wyithe:2004jw}
  Wyithe, J.S.B., Loeb, A., Carilli, C.
  ApJ, 628, 575, 2005,
  \eprint{astro-ph/0411625}
  
\bibitem[{Zahn {et~al.}(2005)Zahn, Zaldarriaga, Hernquist, \&
  McQuinn}]{Zahn:2005fn}
Zahn, O., Zaldarriaga, M., Hernquist, L., \& McQuinn, M. 2005, ApJ, 630, 657,
  \eprint{astro-ph/0503166}

\bibitem[{Zahn {et~al.}(2006) in prep.}]{Zahn:2006fi}
Zahn, O., et. al, in preparation
  

\bibitem[{Zaldarriaga {et~al.}(2004)Zaldarriaga, Furlanetto, \&
  Hernquist}]{Zaldarriaga:2003du}
Zaldarriaga, M., Furlanetto, S.~R., \& Hernquist, L. 2004, ApJ, 608, 622,
  \eprint{astro-ph/0311514}

\end{thebibliography}
\end{document}